\documentclass[authoryear,a4paper,review]{elsarticle}
\usepackage{amsmath,amssymb,natbib,graphicx,hyperref,soul,xparse,setspace,booktabs}
\usepackage[modulo]{lineno}
\usepackage[left=1in,right=1in,top=1.4in,bottom=1.2in]{geometry}

\hypersetup{
pdftitle={Yield Spread Selection in Predicting Recession Probabilities: A Machine Learning Approach},
pdfauthor={Jaehyuk Choi, Desheng Ge, Kyu Ho Kang, Sungbin Sohn},
pdfkeywords={},
colorlinks=true,
linkcolor=red,
citecolor=blue,
urlcolor=blue,
bookmarksnumbered=true,
pdfstartview=}
\NewDocumentCommand{\parencite}{O{}O{}m}{\citep[#1#2]{#3}}
\newcommand{\textcite}[2][]{\citet[#1]{#2}}

\begin{document}

\date{\today}

\begin{frontmatter}
\title{Yield Spread Selection in Predicting Recession Probabilities:\\A Machine Learning Approach}

\author[phbs]{Jaehyuk Choi} \ead{jaehyuk@phbs.pku.edu.cn}
\author[phbs]{Desheng Ge} \ead{gdsuibe@gmail.com}
\author[ku]{Kyu Ho Kang\corref{corrauthor}} \ead{kyuho@korea.ac.kr}
\author[sogangadd]{Sungbin Sohn} \ead{sungbsohn@sogang.ac.kr}

\cortext[corrauthor]{Corresponding author. Tel.: +82 2-3290-5132. Kyu Ho Kang acknowledges financial support by Korea University(K2108911)}
\address[phbs]{Peking University HSBC Business School, University Town, Shenzhen 518055, China}
\address[ku]{Korea University, Department of Economics, Seoul 02841, Republic of Korea}
\address[sogangadd]{Sogang University, School of Economics, Seoul 04107, Republic of Korea}

\begin{abstract}
The literature on using yield curves to forecast recessions customarily uses 10-year--three-month Treasury yield spread without verification on the pair selection. This study investigates whether the predictive ability of spread can be improved by letting a machine learning algorithm identify the best maturity pair and coefficients. Our comprehensive analysis shows that, despite the likelihood gain, the machine learning approach does not significantly improve prediction, owing to the estimation error. This is robust to the forecasting horizon, control variable, sample period, and oversampling of the recession observations. Our finding supports the use of the 10-year--three-month spread.
\end{abstract}

\end{frontmatter}

\noindent \emph{JEL classification}: C52, E32, E43

\noindent \emph{Keywords}: Yield curve, estimation risk, density forecasting, machine learning

\section{Introduction}

\noindent \label{sec:intro} The term spread is a well-established parsimonious and high frequency
indicator of recessions, and it is particularly useful for policy-makers and financial market participants. On average, the yield curve is gently upward
sloping and concave. However, over the past 50 years, recessions have often been preceded by a flattening or even an inversion of the curve. Motivated
by this stylized fact, numerous empirical studies have documented the
ability of the slope of the yield curve or the term spread to predict
future recessions.\footnote{
\citet{stock1989new} and \citet{estrella1991term} find evidence in
support of the predictive power of the slope of the yield curve for
continuous real activity variables. Since then, several works have reported the
predictive ability of the term spread for recessions as well, including \citet{estrella2005yield}, \citet{rudebusch2009forecasting},
\citet{ergungor2016recession}, \citet{engstrom2018near},
\citet{Johansson2018}, and \citet{bauer2018economic}. In particular,
\citet{rudebusch2009forecasting} show that a simple prediction model based
on the term spread outperforms professional forecasters at predicting
recessions three and four quarters ahead.}

The literature on using the yield curve to predict recessions typically
measures the term spread as the difference between the 10-year bond yield and the three-month bond
yield,\footnote{
This term spread is analyzed by
\citet{estrella2005yield}, \citet{estrella2006yield}, and \citet{stekler2017evaluating}. See also
\citet{bauer2018information} for a performance comparison with other
term spreads.} and tends to focus on quantifying the predictive
power of the term spread at a particular forecasting horizon. The
forecasting horizon is usually chosen at four quarters, where the predictive
ability of the term spread is maximized \citep{estrella2006yield,Berge2011}.
Few studies have formally discussed criteria for
selecting a pair of short- and long-term interest rates from a number of bond yields with different maturities. Moreover, using the term spread as a
predictor implies that the coefficients of the
short-term and long-term interest rates are constrained to have the same magnitude with opposite signs.

Government bond yields are determined by the sum of the market expectation
and the risk premium.
Given that the market expectation component is the expected path of future short-term rates and the policy rates are pro-cyclical, if the risk premium or its variation is small, the term spread contains much predictive information about future business cycles and future policy rates.
However, the portion of the market expectation component in a bond yield differs across the maturity. In addition, liquidity risk premiums are strongly time varying, particularly in the short-term Treasury bill markets~\citep{Goyenko2011}. Therefore, the predictive ability of the term spread can be sensitive to the maturity combination. For the same reason, the absolute regression effects
of the interest rates on the recession probability are not necessarily the
same, and the short- and long-term interest rates may have separate effects
from the spread.

The question we address here is whether relaxing the restrictions on the
fixed maturity pair and the coefficients can improve the predictive ability of the bond yields. This question is particularly important from a statistical point of view. If the answer is yes, then the choice of maturity pair should be included in the prediction procedure, and the short- and long-term yields should be used as separate predictors. 
However, if the answer is no, it implies the substantial estimation error, and the conventional use of the 10-year--three-month Treasury yield spread is justified.

To answer the question, we use a machine learning (ML) framework to search for the best maturity combination and the coefficients of the interest rates
simultaneously. Specifically, we select two maturities from among the nine bond yield series based on a logistic regression with $L_{1}$
regularization for multiple forecasting horizons. Next, we validate the
classification algorithm by comparing the out-of-sample prediction against
the benchmark spread (i.e., the 10-year--three-month term spread).

Our two key findings are based on US data for the period June 1961 to July 2020. First, the optimal maturity pairs for most horizons vary from the (10 year, three month) pair. For instance, for the one-quarter-ahead recession prediction, (10 year, six month) is selected by the ML algorithm. The 20-year--one-year spread performs best for horizons of seven and eight quarters. In addition, the absolute effect of the short-term yield is found to be larger than that of the long-term yield. Second, and more importantly, we find that the prediction gain from the maturity pair optimization or separation of the regression effects is not statistically significant. In particular, the benchmark spread provides better forecasts than those of the ML approach for short- and medium-term horizons. These findings are robust, even when controlling for the leading business cycle index. The poor performance of the proposed approach indicates that the efficiency loss from the estimation risk dominates the likelihood gain.
In summary, based on a comprehensive empirical analysis, we provide new and interesting evidence justifying the conventional use of the 10-year--three-month term spread.

This paper contributes to the growing literature on the application of ML to economic research and the controversy over its effectiveness. While several studies~\citep{gogas2015yield,dopke2017predicting,hall2018machine} claim outperformance of the ML method, \citet{puglia2020ml} report that the prediction gain of ML, in recession forecast at least, is reversed when they employ a more conservative cross validation method to avoid possible overfitting. This study provides evidence in support of \citet{puglia2020ml}.

Stressing what we do not claim is also important. First of all, the disappointing performance of ML should not be interpreted as an evidence for rejecting ML in general.
We use a particular regularization estimator and ML offers a much broader range of alternative prediction procedures. We do not claim that any alternative ML procedures are less
adequate for the problem.
Moreover, given that our interest is centered on the pair of short- and long-term yields, we restrictively use ML to improve the prediction of a spread, rather than the whole yield curve.

The remainder of this paper is organized as follows. In Section~\ref{sec:ml}, we discuss our ML algorithm and data. Section~\ref{sec:main} presents our findings, and Section~\ref{sec:conc} concludes the paper.

\bigskip

\section{Machine Learning Algorithm and Data}

\label{sec:ml} \noindent

\subsection{Logistic Regression with $L_{1}$ Regularization}

\label{ssec:LRL1} \noindent We let $y_{t}$ denote the binary recession
indicator at month $t$ (one if a recession, zero if not). The $k$-month-ahead recession probability $\hat{y}_{t+k}$ at month $t$ is predicted as
\begin{equation*}
\hat{y}_{t+k}=\text{Prob}(y_{t+k}=1|\mathbf{x}_{t})=\phi (-\beta_{0}-
\boldsymbol{\beta}^{T}\mathbf{x}_{t}),
\end{equation*}
where $\phi (z)=(1+e^{-z})^{-1}$ is the logistic function, $\mathbf{x}_{t}$ is the Treasury yield vector of $p$ different maturities at month $t$, $\boldsymbol{\beta}$ is the
corresponding coefficient vector, and $\beta_{0}$ is the intercept. Note
that the sign of $\beta_{0}+\boldsymbol{\beta}^{T}\mathbf{x}_{t}$ is
deliberately negated inside $\phi(\cdot)$ so that
the recession probability increases when the linear combination of the
yields decreases. This ensures consistency with the
stylized fact that a recession is typically followed by a negative term
spread, defined as the long-term yield minus the short-term yield.

Our model differs from a traditional logistic regression in that large coefficient values are penalized during the maximization of the likelihood. This measure, called \textit{regularization} in machine learning literature, prevents overfitting in the prediction problems with small sample size or large number of predictors. Specifically, we find $\boldsymbol{\beta}$ and $\beta_{0}$ that minimize
the cost function,
\begin{equation}
J(\beta_{0},\boldsymbol{\beta})=-\log L(\beta_{0},\boldsymbol{\beta}
)+\lambda \,\lVert \boldsymbol{\beta}\rVert_{1}\quad (\lambda \geq 0),
\label{eq:ObjFunc}
\end{equation}
where $\lVert \boldsymbol{\beta}\rVert_{1}=|\beta_{1}|+\cdots +|\beta
_{p}|$ is the $L_{1}$ norm of the coefficient vector, and $\log L$ is the log likelihood over the training period $\mathcal{T}$,
\begin{equation}
\log L(\beta_{0},\boldsymbol{\beta})=\sum_{t+k\in \mathcal{T}}\left(
y_{t+k}\ln (\hat{y}_{t+k})+(1-y_{t+k})\ln (1-\hat{y}_{t+k})\right).
\label{eq:loglh}
\end{equation}
The regression for continuous variables with the same $L_{1}$ regularization is well known as the least absolute shrinkage and selection operator (LASSO)~\citep{hastie2009ESL}. The LASSO differs from the ridge regression, which
instead uses the $L_{2}$ norm $\lVert \boldsymbol{\beta}\rVert_{2}=\beta
_{1}^{2}+\cdots +\beta_{p}^{2}$ for penalty. The use of $L_{1}$ or $L_{2}$
regularization tends to make the coefficients smaller in magnitude as the
regularization strength $\lambda $ increases, hence the shrinkage regression.
The shrinkage regression method has recently been adopted in economic
forecasting. For example, \citet{hall2018machine} uses the elastic-net
regularization~\citep{zou2005elastic}, where both $L_{1}$ and $L_{2}$
penalties are used for regularization; see \citet{kim2018mining} for the
effectiveness of the shrinkage method in predicting various macroeconomic
variables.

Although the $L_1$ penalty makes the optimization more difficult than when using the quadratic $L_2$ penalty, it is attractive because
it forces certain coefficients to be set to zero; see the graphical
interpretation in \citet[Figure 3.11]{hastie2009ESL}. Therefore, it
effectively simplifies the model by using only a small subset of the input
variables, thus performing a feature selection. In contrast, the $L_2$
regularization does not set the coefficients to zero, although it does shrink the magnitudes.

The logit model with $L_1$ regularization is a natural choice of
algorithm for this study because we aim to simultaneously extract two maturities and find their coefficients from the term structure, without any prior knowledge. We do not consider more complicated ML algorithms that make full use of the rates from all maturities. Indeed, previous studies have adopted methods such as support vector machines~\citep{gogas2015yield} and boosted regression trees~\citep{dopke2017predicting}. Their strategy for recession prediction is conducting a variable selection among a large number of macroeconomic predictors in a nonlinear fashion or using the entire yield curve information, while our study aims to search the best maturity pair of term spread.

Note the following with regard to implementation. Because the coefficient
magnitude is in the objective function, equation~\eqref{eq:ObjFunc}, the
optimization result depends on the scale of the input variables. Therefore,
each feature variable is normalized using a z-score before the optimization~\citep{hastie2009ESL}. In the remainder of the paper, however, the
coefficients $(\beta_{0},\boldsymbol{\beta})$ are reported in the
original scale, unless stated otherwise. We use the regularized
logistic regression in the \texttt{scikit-learn} Python package of \citet{pedregosa2011sklearn} to solve $\beta_{0}$ and $\boldsymbol{\beta}$, given $\lambda $.\footnote{The class can be downloaded from
scikit-learn.org} In searching for $\lambda $, we geometrically increase the regularization strength as $\lambda =2^{\,k/10}$ from a large negative integer value of $k$. Then, we report the first value of $\lambda $ at which the number of
nonzero coefficients in $\boldsymbol{\beta}$ becomes the desired
number (i.e., two).

\subsection{Competing Specifications} \label{ssec:specs}

\noindent Suppose that the pair of Treasury yields at month $t$
extracted from the $L_{1}$ regularization are denoted by ($x_{i,t},$ $
x_{j,t} $), and are dependent on the forecasting horizon. The resulting
predictive model is given by
\begin{gather*}
\mathcal{M}(\mathit{generalized\ spread\ of\ the\ ML\ pair}) \\
\hat{y}_{t+k}=\phi (-\beta_{0}-\beta_{i}x_{i,t}-\beta_{j}x_{j,t}).
\end{gather*}
We refer to the linear combination of the yields and a constant, $\beta
_{0}+\beta_{i}x_{i,t}+\beta_{j}x_{j,t}$, as a \textit{generalized spread}
of the ML pair. To evaluate the prediction performance of the generalized
spread, we compare the proposed model with its three nested logit models.
The first alternative model uses the term spread measured by the simple
difference of the ML pair (hereafter, the \textit{simple spread} of the ML pair). That is, the model is given by
\begin{gather*}
\mathcal{M}(\mathit{simple\ spread\ of\ the\ ML\ pair}) \\
\hat{y}_{t+k}=\phi (-\beta_{0}-\beta_{i}(x_{i,t}-x_{j,t})),
\end{gather*}
where the absolute regression effects of $x_{i,t}$ and $x_{j,t}$ are
constrained to be the same (i.e., $\beta_{i}=-\beta_{j}$).

In the second alternative model, the conventional yield pair is used without the
coefficient restriction, as follows:
\begin{gather*}
\mathcal{M}(\mathit{generalized\ spread\ of\ the\ conventional\ pair}) \\
\hat{y}_{t+k}=\phi (-\beta_{0}-\beta_{i}l_{t}-\beta_{j}s_{t}),
\end{gather*}
where $l_{t}$ and $s_{t}$ are the 10-year and three-month yields, respectively, at month $t$. The third alternative model is the benchmark and the most restricted version of the proposed model, in which the simple spread of the
conventional pair is used as a predictor:
\begin{gather*}
\mathcal{M}(\mathit{simple\ spread\ of\ the\ conventional\ pair}) \\
\hat{y}_{t+k}=\phi (-\beta_{0}-\beta_{i}(l_{t}-s_{t})).
\end{gather*}

By comparing the proposed model with these nested frameworks, we
separately identify the importance of the maturity pair selection and that of the
coefficient restriction. Note that the model parameters are searched in order to maximize the log likelihood in equation \eqref{eq:loglh} without
regularization in the nested models.

\subsection{Data}

\noindent We describe the data used in the analysis. For the binary state of a recession, $y_{t}$, we use the monthly periods of recessions defined by the National Bureau of Economic Research (NBER). For the yield curve $\mathbf{x}_{t}$, we use the monthly averaged constant maturities time series of the Treasury yields from the H.15 page of the Federal Reserve website. Our sample period is June 1961 to July 2020. For this period, we use three- and six-month and one-, two-, three-, five-, seven-, 10-, and 20-year yields. Because not all series are available in the early part of the sample period, alternative sources are used to fill in the missing data. The three- and six-month Treasury bill rates until August 1981 are taken as the secondary market rates from the same website. Because they are recorded on a discount basis, we convert them to a bond-equivalent basis, as in \citet{estrella2005yield}. The two- and seven-year yields until May 1976 and June 1969, respectively, are obtained from the zero-coupon Treasury yield curve by \citet{gurkaynak2006treasury,gurkaynak2007treasury}. We split the overall period into training (in-sample) and test (out-of-sample) periods at the end of 1995.

\section{Results}

\label{sec:main} \noindent This section reports the results of the best pair
selection for a wide range of forecasting horizons, including its associated coefficients, and discusses the implications of these results. As a baseline examination, we use the period June 1961 to December 1995 for training (pair selection and coefficient estimation), and the period since 1996 for testing (out-of-sample evaluation). Then, we check the robustness of the findings by trying alternative training/test periods and handling the problems of imbalanced classification and missing variables.

\begin{figure}[tbp]
\caption{The path of the coefficients in forecasting using the 12-month horizon
as a function of the $L_1$ regularization strength $\protect\lambda$ with a 12-month forecasting horizon. The black dotted line is the value of $\protect
\lambda=0.3536$, where only two variables (i.e., the seven-year and three-month yields) are left. The training period is June 1961 to December 1995.}
\label{fig:CoefPath}\centering
\includegraphics[width=0.7\linewidth]{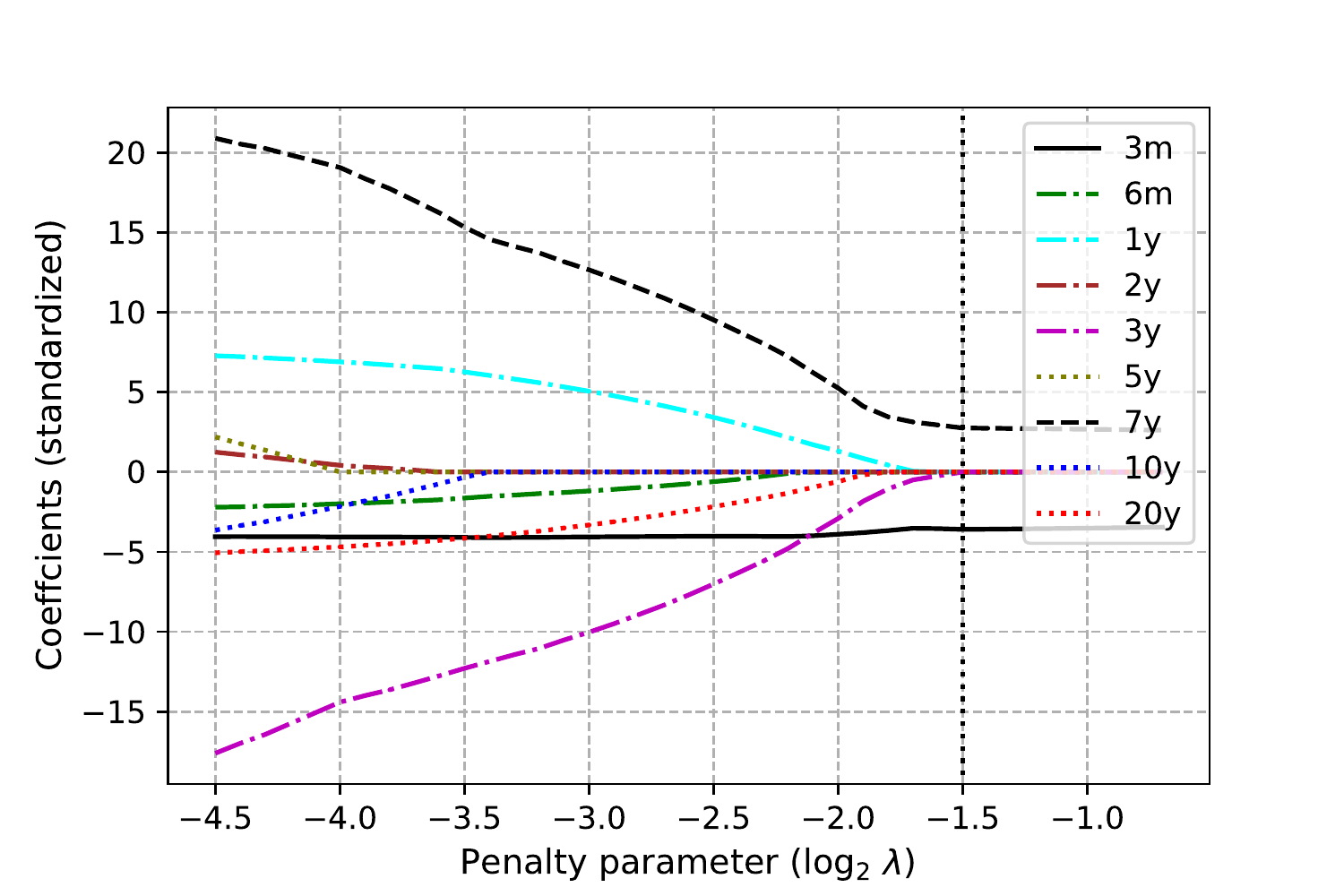}
\end{figure}

\subsection{Pair and Coefficient Selection Using Machine Learning}

\noindent Figure~\ref{fig:CoefPath} visually illustrates the best pair-selection process described in Section~\ref{ssec:LRL1}. As we increase the
magnitude of the penalty parameter, $\lambda $, in equation (\ref{eq:ObjFunc}), the coefficients of the features with little
contribution to the likelihood maximization are set to zero. We continue
increasing the penalty strength until only two yields survive. In the case of
the 12-month-ahead recession prediction, three-month and seven-year yields
survive as the pair that best explains the recession from June 1961 to December 1995.

Panel A in Table~\ref{tab:MainResult95} summarizes the results of the
ML analyses for the various forecasting horizons. There are
two main findings. First, we find that for every forecasting horizon, one
long-term and one short-term yields are selected as the best pair for
recession prediction, and their coefficients have positive
and negative signs, respectively. This confirms the use of the term spread in the past literature. However, unlike the conventional approach, the best prediction performance is achieved with a pair other than the 10-year and three-month yields. Specifically, when the forecasting horizon is relatively short (three or six months), the Treasury yields with 10-year and six-month maturities are the best predictors of a recession. For long horizons (18, 21, 24 months), the longest available term (20-year) yield and a short-term yield (three-month, six-month, or one-year, depending on the forecasting horizon) work as the best pair. These findings are roughly consistent with the notion that longer-term yields reflect information about a relatively more distant future.

\begin{table}[p]
\caption{\textbf{Pair and coefficient selection from the training
period, 1961--1995} Panel A presents the results for the best pair and coefficients
selected by machine leaning for several forecasting horizons. Panel B
presents the recession prediction performance for the simple term spread of
the pair selected in Panel A. Panel C (D) presents the performance for the
generalized (simple) term spread of the conventional 10-year and three-month
pair. $\protect\lambda$ is the strength of the $L_1$ penalty, $\text{AUC}_\text{train}$ ($\text{AUC}_\text{test}$) is the area under the ROC in the training (test) period, log L (log PPL) is the average log likelihood (PPL) in the
training (test) period, and the EBF is the ratio of the
PPL of an alternative model to that of the benchmark model. The longest
maturity in the sample is 20 years.}
\label{tab:MainResult95}
\begin{center} \small
\begin{tabular}{ccccccccc}
\hline\hline
Horizon & Pair & $\boldsymbol{\beta}$ & $\lambda$ & $\text{AUC}_\text{train}$
& $\text{AUC}_\text{test}$ & log L & log PPL & EBF \\ \hline
\multicolumn{9}{l}{\textit{Panel A. Generalized spread of the ML pair}} \\
3 & (10y, 6m) & (0.453, -0.790) & 0.871 & 0.902 & 0.529 & -0.282 & -0.459 &
0.912 \\
6 & (10y, 6m) & (0.773, -1.069) & 0.933 & 0.937 & 0.644 & -0.255 & -0.413 &
0.931 \\
9 & (10y, 3m) & (0.926, -1.191) & 0.933 & 0.939 & 0.766 & -0.246 & -0.334 &
0.956 \\
12 & (7y, 3m) & (1.039, -1.231) & 0.354 & 0.919 & 0.850 & -0.270 & -0.277 &
0.981 \\
15 & (7y, 3m) & (0.893, -1.010) & 0.500 & 0.868 & 0.890 & -0.304 & -0.258 &
0.990 \\
18 & (20y, 3m) & (0.428, -0.538) & 2.639 & 0.806 & 0.899 & -0.343 & -0.271 &
0.985 \\
21 & (20y, 6m) & (0.402, -0.464) & 1.741 & 0.744 & 0.898 & -0.369 & -0.276 &
1.000 \\
24 & (20y, 1y) & (0.383, -0.409) & 1.231 & 0.671 & 0.892 & -0.391 & -0.283 &
1.016 \\
\multicolumn{9}{l}{\textit{Panel B. Simple spread of the ML pair}} \\
3 & (10y, 6m) & (0.956, -0.956) &  & 0.810 & 0.523 & -0.319 & -0.380 & 0.987
\\
6 & (10y, 6m) & (1.274, -1.274) &  & 0.879 & 0.657 & -0.279 & -0.353 & 0.989
\\
9 & (10y, 3m) & (1.396, -1.396) &  & 0.901 & 0.780 & -0.263 & -0.289 & 1.000
\\
12 & (7y, 3m) & (1.342, -1.342) &  & 0.892 & 0.861 & -0.280 & -0.250 & 1.008
\\
15 & (7y, 3m) & (1.119, -1.119) &  & 0.857 & 0.898 & -0.308 & -0.247 & 1.001
\\
18 & (20y, 3m) & (0.759, -0.759) &  & 0.791 & 0.916 & -0.341 & -0.252 & 1.004
\\
21 & (20y, 6m) & (0.599, -0.599) &  & 0.733 & 0.907 & -0.368 & -0.265 & 1.011
\\
24 & (20y, 1y) & (0.518, -0.518) &  & 0.669 & 0.903 & -0.390 & -0.273 & 1.026
\\
\multicolumn{9}{l}{\textit{Panel C. Generalized spread of the conventional pair}}
\\
3 & (10y, 3m) & (0.407, -0.769) &  & 0.901 & 0.519 & -0.288 & -0.458 & 0.913
\\
6 & (10y, 3m) & (0.762, -1.094) &  & 0.940 & 0.630 & -0.258 & -0.416 & 0.929
\\
9 & (10y, 3m) & (1.031, -1.295) &  & 0.938 & 0.767 & -0.245 & -0.338 & 0.952
\\
12 & (10y, 3m) & (0.983, -1.162) &  & 0.915 & 0.836 & -0.271 & -0.284 & 0.974
\\
15 & (10y, 3m) & (0.841, -0.951) &  & 0.858 & 0.881 & -0.308 & -0.260 & 0.988
\\
18 & (10y, 3m) & (0.646, -0.738) &  & 0.796 & 0.893 & -0.341 & -0.263 & 0.993
\\
21 & (10y, 3m) & (0.438, -0.510) &  & 0.709 & 0.880 & -0.376 & -0.277 & 0.999
\\
24 & (10y, 3m) & (0.245, -0.292) &  & 0.619 & 0.877 & -0.403 & -0.297 & 1.002
\\
\multicolumn{9}{l}{\textit{Panel D. Simple spread of the conventional pair}} \\
3 & (10y, 3m) & (0.823, -0.823) &  & 0.794 & 0.521 & -0.333 & -0.367 & 1.000
\\
6 & (10y, 3m) & (1.158, -1.158) &  & 0.871 & 0.654 & -0.290 & -0.342 & 1.000
\\
9 & (10y, 3m) & (1.396, -1.396) &  & 0.901 & 0.780 & -0.263 & -0.289 & 1.000
\\
12 & (10y, 3m) & (1.258, -1.258) &  & 0.894 & 0.846 & -0.280 & -0.258 & 1.000
\\
15 & (10y, 3m) & (1.018, -1.018) &  & 0.850 & 0.890 & -0.312 & -0.248 & 1.000
\\
18 & (10y, 3m) & (0.786, -0.786) &  & 0.783 & 0.901 & -0.344 & -0.256 & 1.000
\\
21 & (10y, 3m) & (0.541, -0.541) &  & 0.697 & 0.890 & -0.378 & -0.276 & 1.000
\\
24 & (10y, 3m) & (0.305, -0.305) &  & 0.610 & 0.892 & -0.404 & -0.299 & 1.000
\\ \hline
\end{tabular}
\end{center}
\end{table}

Second, the optimal coefficients of the generalized term spread are close to, but different from $(1,-1)$. In predicting the 12-month-ahead recession
using the training period up to 1995, the coefficients for the seven-year and three-month yields are 1.039 and $-1.231$, respectively. For all forecasting horizons, the coefficient of the short-term yield is bigger in magnitude than that of the associated long-term yield, suggesting that the short-term yield plays a greater role in recession prediction than implied by a simple term spread. Interestingly, we find that the larger absolute effect of the short-term yields is more pronounced, particularly when the forecasting horizon is shorter. For example, the absolute coefficient ratio of shorter- and longer-term yields is 1.744 $\left( =\left\vert \frac{-0.790}{0.453}
\right\vert \right) $ in three-month-ahead forecasting, but this decreases to 1.382 $\left( =\left\vert \frac{-1.069}{0.733}\right\vert \right) $, 1.184 $
\left( =\left\vert \frac{-1.231}{1.039}\right\vert \right)$, and 1.069 $
\left( =\left\vert \frac{-0.409}{0.383}\right\vert \right)$ for six-, 12- and 24-month-ahead forecasting, respectively. This pattern implies that the
ML approach may improve the simple term spread, particularly
for shorter-horizon recession predictions.

\begin{figure}[tbp]
\caption{\textbf{The generalized term spread and the implied recession
probability 12 months ahead}. The shaded vertical bars indicate the
recession periods defined by the NBER. The dotted black line divides the
training (in-sample) and test (out-of-sample) periods; the training period
is June 1961 to December 1995, and the test period is until July 2020.
Note that the train--test split in the upper panel is 12 months behind (i.e., December 1994), owing to the forecasting horizon.}
\label{fig:GSpread_RecProb}\centering \vspace{2ex}
\includegraphics[width=0.8\linewidth]{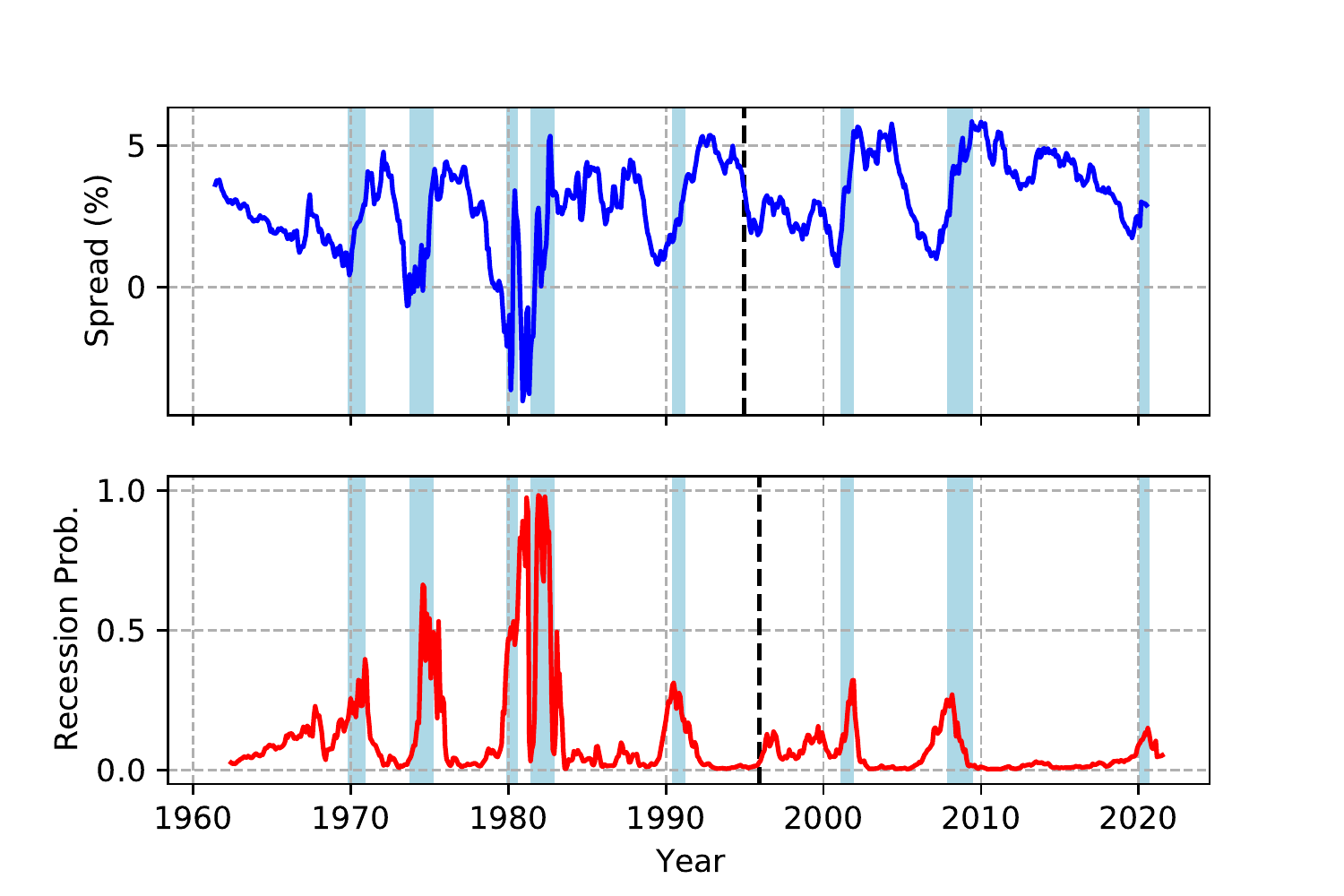}
\end{figure}

Figure~\ref{fig:GSpread_RecProb} depicts the generalized term spread (upper
panel) and the implied 12-month-ahead recession probability (lower panel).
The shaded vertical bar indicates the recession periods, and the dotted black line divides the training (in-sample) and test (out-of-sample) periods. We observe clear spikes in the probability during, or ahead of, recessions. Interestingly, the most recent recession (starting March 2020) is quite precisely detected by the generalized term spread, despite the
fact that we use the sample only until 1995 for the parameter estimation, and that the recession was officially announced in June 2020. This figure
provides a visual validation for the generalized term spread in recession prediction.

\subsection{Prediction Performance Evaluation}

\noindent The key question we address in this study is whether the
generalized term spread based on the machine leaning approach outperforms
the simple term spread of the conventional pair. To this end,
we compare the out-of-sample recession prediction performance of the four
competing models; the results can be found in Table~\ref{tab:MainResult95}.

The predictive recession probability accuracy is measured by the log
posterior predictive likelihood (PPL). The empirical Bayes factor (EBF)
is the ratio of the PPL of an alternative model to that of the benchmark model.
An EBF larger than one indicates stronger support for the alternative model than the benchmark model by the data. Table \ref{tab:MainResult95} shows that, for all forecasting horizons, the proposed model, $\mathcal{M}$(\textit{generalized\ spread\ of\
the\ ML\ pair)}, is not preferred to its nested competing models in terms of the log PPL. For the horizons of three, six, and nine months, the benchmark, $\mathcal{M}$(\textit{simple\ spread\ of\ the\ conventional\ pair}), provides the best forecasts.
Although $\mathcal{M}$(\textit{simple\ spread\ of\ the\ ML\ pair}
)\ outperforms the benchmark for the other horizons, it is not statistically significant, because the largest EBF is at most 1.026. All EBFs are much less than $\sqrt{10}$, regardless of the horizon. Based on Jeffreys' criterion, the evidence that any alternative model is more supported by the data than the benchmark model is very weak. Therefore, the prediction
gain from choosing the maturity pair or relaxing the coefficient restriction
is not substantial.\footnote{We also examine the relative mean squared error (RM) as an alternative forecast evaluation criterion as in \citet{Mincer1969} and \citet{racicot2007forecasting}, and reach the consistent conclusion. Specifically, we find that RMs are mostly greater than 1 and the smallest value is 0.964, meaning that the forecast error of a considered model is not much smaller than that of the benchmark model. The results are available upon request.}

The poor out-of-sample prediction performance of the proposed model seems to arise from the inefficiency in the coefficient estimation, as pointed out in the equity return prediction literature
\citep{welch2008equitypred,demiguel2009naive}. The pair selection itself,
which is relatively less subject to the estimation error, could still be
conducive to improving the recession prediction. We can easily test this
conjecture by comparing the models $\mathcal{M}$(\textit{simple\ spread\ of\
the\ ML\ pair}) and $\mathcal{M}$(\textit{generalized\ spread\ of\ the\
conventional\ pair}). Given that the EBF measures the prediction performance of an alternative model relative to the benchmark, the reciprocal of the EBFs of $\mathcal{M}$(\textit{simple\ spread\ of\ the\ ML\ pair}) presents
the inefficiency from the pair selection. Similarly, the reciprocal of the
EBFs of $\mathcal{M}$(\textit{generalized\ spread\ of\ the\ conventional\
pair}) quantifies the coefficient estimation risk. The EBFs from the simple\
spread\ of\ the\ ML\ pair are larger than or equal to those from the
generalized\ spread\ of\ the\ conventional\ pair, regardless of the horizon. As a result, the inefficiency is attributed more to the coefficient estimation than it is to the pair selection.

\begin{figure}[tbp]
\caption{\textbf{The receiver operating characteristic (ROC) curves for
several forecasting horizons} The lines A, B, C, and D indicate the results for the generalized spread of the ML pair, the simple
spread of ML pair, the generalized spread of the conventional pair, and the simple spread of the conventional pair, respectively. The area under the ROC curve is given in parentheses. The ROC curves are evaluated from the test period, where the
training period is 1961--1995.}
\label{fig:ROC}
\begin{center}
\includegraphics[width=0.4\linewidth]{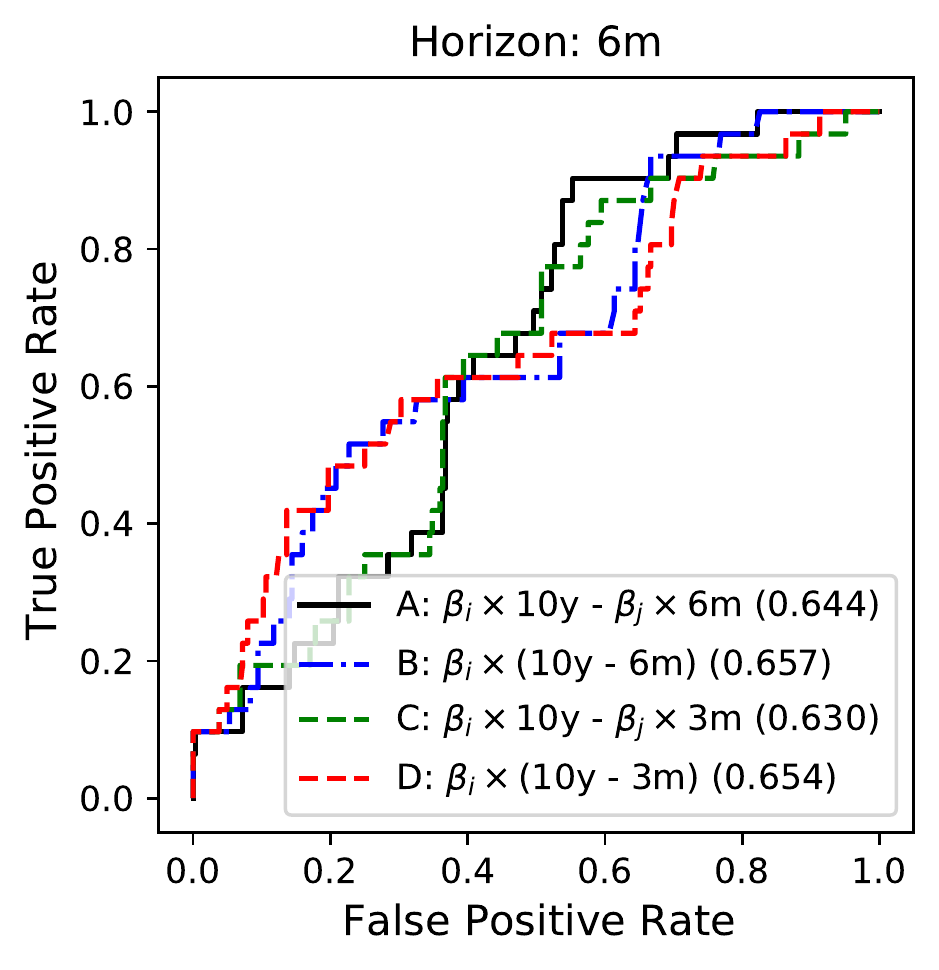}
\includegraphics[width=0.4\linewidth]{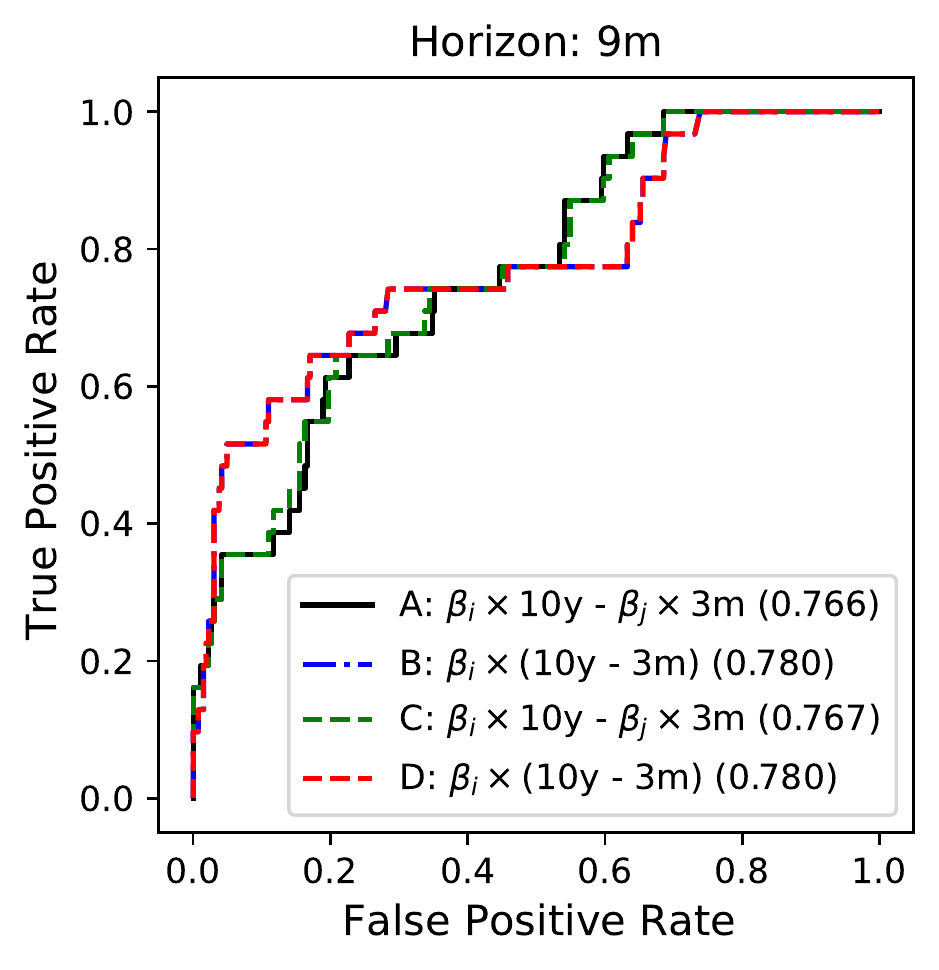}\\[0pt]
\includegraphics[width=0.4\linewidth]{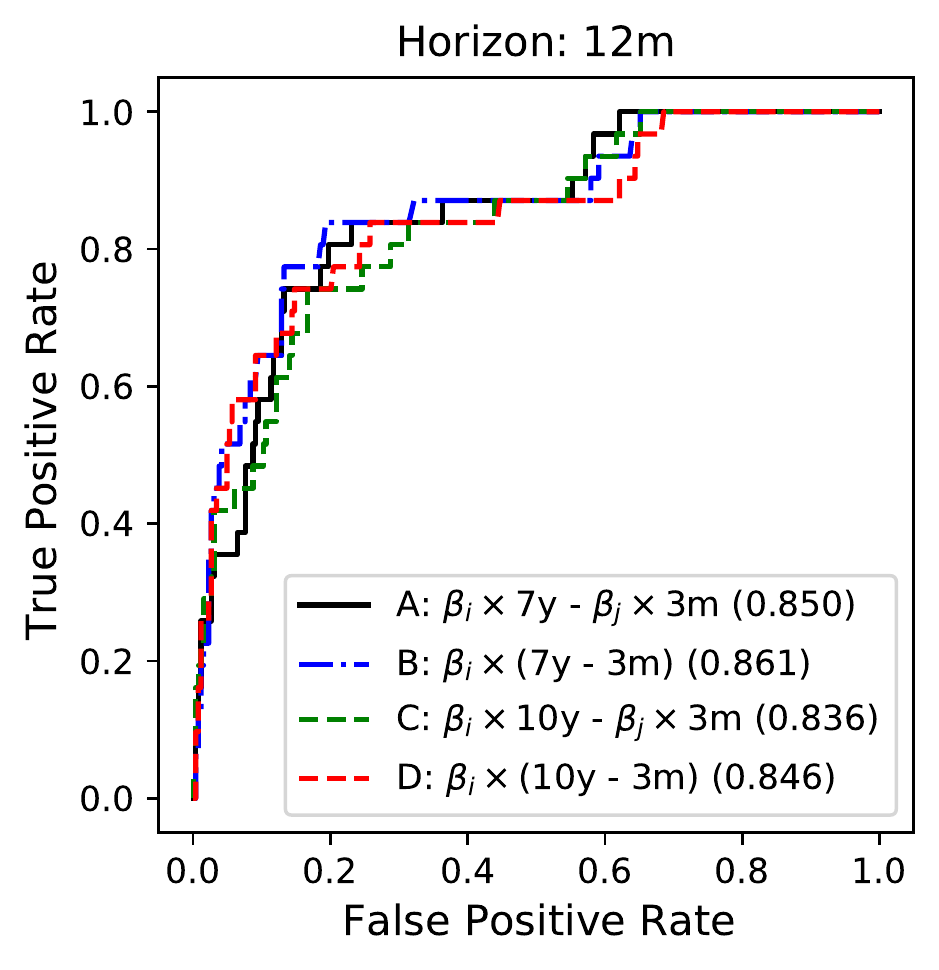}
\includegraphics[width=0.4\linewidth]{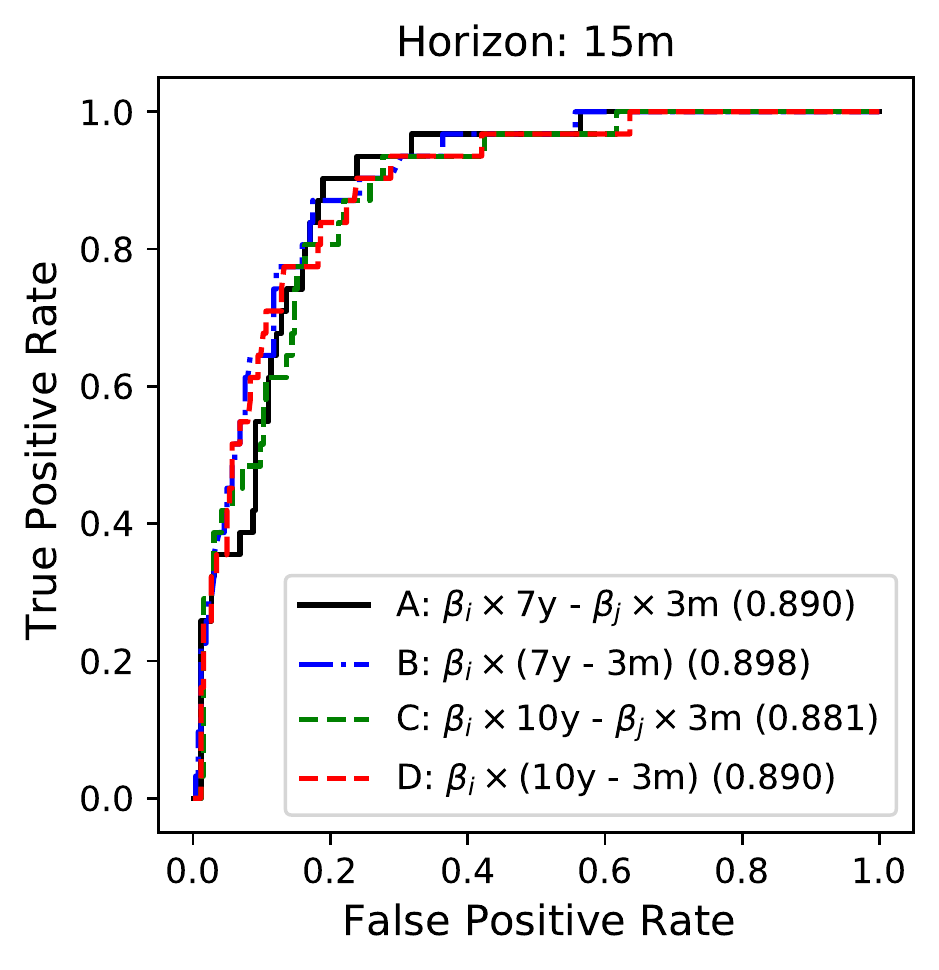}\\[0pt]
\includegraphics[width=0.4\linewidth]{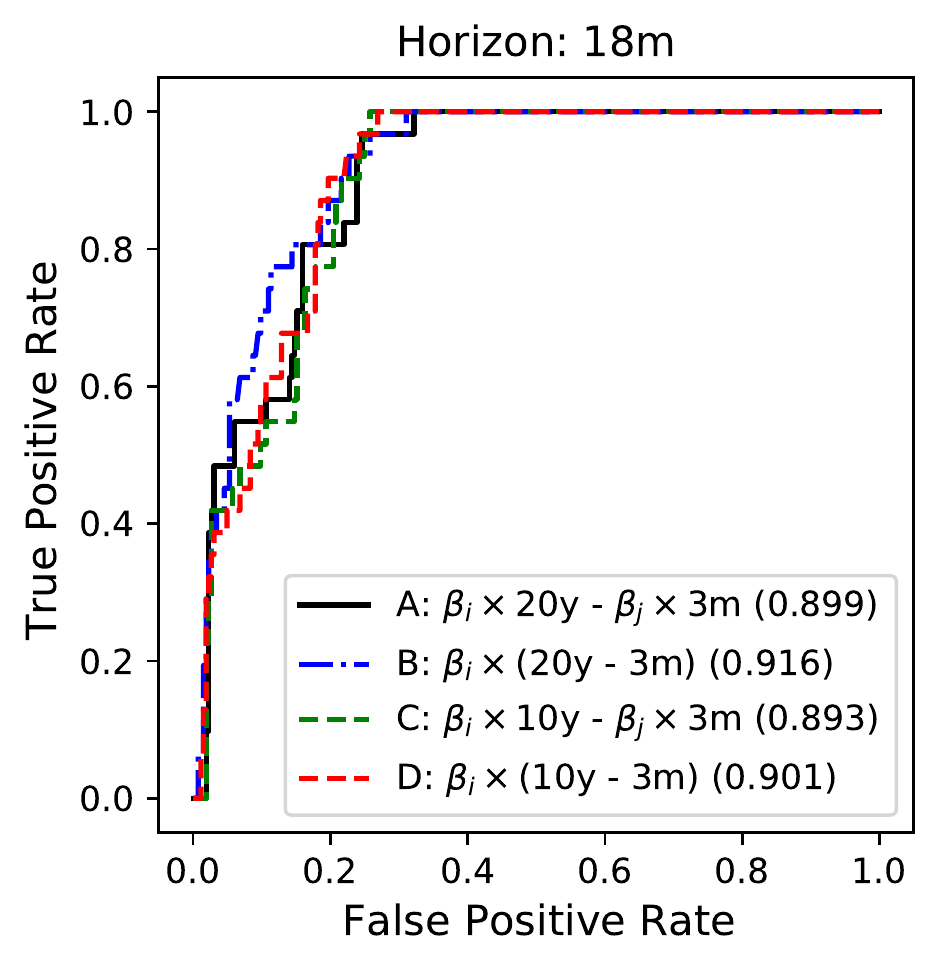}
\includegraphics[width=0.4\linewidth]{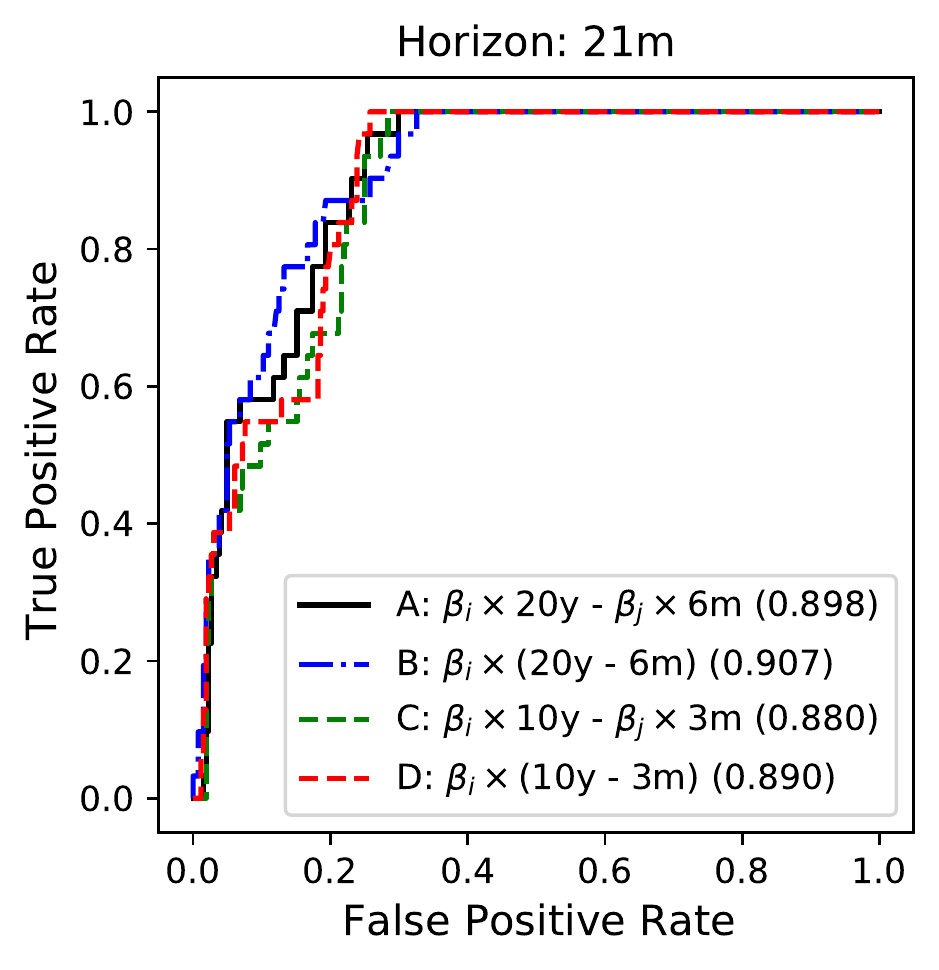}
\end{center}
\end{figure}

We evaluate the area under the ROC curve (hereafter, ROC-AUC)
as a supplementary performance evaluation to the log PPL. The ROC curve
presents a collection of the (false positive rate, true positive rate)
coordinates for various decision thresholds between zero and one. The true
positive rate is the ratio of correct predictions among real recessions
(i.e., related to type-II error). The false positive rate is the ratio
of incorrect predictions among real non-recessions (i.e., related to
type-I error). Similar to the log PPL, the ROC-AUC captures the predictive power of the model without any specific decision threshold; the value is one for a perfect model, and 0.5 for a random guess.
The ROC-AUC is an established performance measure in ML~\citep{bradley1997use}, and has recently been used in the context of recession prediction by \citet{liu2016what}, \citet{stekler2017evaluating}, \citet{bauer2018information} and \citet{tsang2019revisiting}.

Figure~\ref{fig:ROC} depicts the ROCs for six forecasting
horizons, and Table~\ref{tab:MainResult95} reports the ROC-AUCs. These results provide evidence in favor of $\mathcal{M}$(\textit{simple\ spread\ of\ the\
ML\ pair}) in terms of ROC-AUC. The simple spread of the ML pair in the test period exhibits the largest ROC-AUC in most forecasting horizons. Nevertheless, the difference between the ROC-AUCs of $\mathcal{M}$(\textit{simple\ spread\ of\ the\ ML\ pair}) and the benchmark is not substantial in any of the horizons.

Although not intended, our findings validate the widely used, but seemingly
ad hoc conventional term spread. Recall that we attempt to find the best
yield pair for recession probability prediction without imposing any
restrictions. The ML approach does not base its results on
economic theory or academic norms, but only on the best (in-sample) prediction performance. However, the findings support the use of
the conventional 10-year and three-month term spread surprisingly well, in that the pairs from the ML similarly consist of one long- and one short-term yields, and that the coefficients have the opposite signs with similar magnitudes in the absolute sense. Although the ML approach chooses a slightly different pair from the (10-year, three-month) combination and the coefficient ratio modestly deviates from one, the resultant out-of-sample prediction performance is not distinguishable from that of the conventional term spread.

\subsection{Robustness Checks}

\noindent In this subsection, we conduct various robustness checks to ensure that our findings do not result from a specific choice of training/test
samples or oversampling and missing variable problems.

\subsubsection{Training and Test Periods}

First, we try alternative training and test periods. Tables~\ref{tab:MainResult05} and \ref{tab:MainResult15} show the results when the
training period extends to 2005 and 2015, respectively. Accordingly, there are
fewer recession events in the alternative test periods. Although the
specific choice of the best pair varies slightly, the three key findings
remain unchanged: (i) the pair chosen using ML consists of one short-term
and one long-term yields, with coefficients of opposite signs; (ii) $
\mathcal{M}$(\textit{simple\ spread\ of\ the\ ML\ pair}) seems to be the
best in terms of the log PPL, particularly for a longer forecasting horizon;
and (iii) the maturity pair selection or coefficient estimation separating
the effects of the short- and long-term yields does not improve the
predictive accuracy significantly. In particular, the
largest EBFs in Tables~\ref{tab:MainResult05} and \ref{tab:MainResult15} are
1.097 and 1.114, respectively. These EBFs imply that the model weights on
the best alternative model are less than 0.53 in an empirical Bayesian model
averaging framework. Consequently, the contribution of the ML
approach to the improvement of the log PPL is at most marginal, if any.

\begin{table}[p]
\caption{\textbf{Pair and coefficient selection from the training
period, 1961--2005} Panel A presents the results for the best pair and coefficients
selected by machine leaning for several forecasting horizons. Panel B
presents the recession prediction performance for the simple term spread of
the pair selected in Panel A. Panel C (D) presents the performance for the
generalized (simple) term spread of the conventional 10-year and three-month
pair. $\protect\lambda$ is the strength of the $L_1$ penalty, $\text{AUC}_\text{train}$ ($\text{AUC}_\text{test}$) is the area under the ROC in the training (test) period, log L (log PPL) is the average log likelihood (PPL) in the
training (test) period, and the EBF is the ratio of the
PPL of an alternative model to that of the benchmark model. The longest
maturity in the sample is 20 years.}
\label{tab:MainResult05}
\begin{center} \small
\begin{tabular}{ccccccccc}
\hline\hline
Horizon & Pair & $\boldsymbol{\beta}$ & $\lambda$ & $\text{AUC}_\text{train}$
& $\text{AUC}_\text{test}$ & log L & log PPL & EBF \\ \hline
\multicolumn{9}{l}{\textit{Panel A. Generalized spread of the ML pair}} \\
3 & (10y, 6m) & (0.414, -0.713) & 0.933 & 0.851 & 0.605 & -0.278 & -0.551 &
0.904 \\
6 & (10y, 3m) & (0.597, -0.901) & 2.297 & 0.925 & 0.655 & -0.248 & -0.518 &
0.945 \\
9 & (10y, 3m) & (1.174, -1.426) & 0.616 & 0.946 & 0.749 & -0.220 & -0.474 &
0.944 \\
12 & (20y, 3m) & (0.486, -0.695) & 5.278 & 0.924 & 0.844 & -0.262 & -0.381 &
0.953 \\
15 & (20y, 3m) & (0.659, -0.789) & 2.144 & 0.874 & 0.913 & -0.282 & -0.324 &
0.955 \\
18 & (20y, 6m) & (0.573, -0.663) & 1.741 & 0.822 & 0.967 & -0.310 & -0.303 &
0.970 \\
21 & (20y, 1y) & (0.636, -0.685) & 0.536 & 0.759 & 0.982 & -0.334 & -0.293 &
1.000 \\
24 & (20y, 1y) & (0.397, -0.434) & 1.414 & 0.679 & 0.985 & -0.356 & -0.310 &
1.013 \\
\multicolumn{9}{l}{\textit{Panel B. Simple spread of the ML pair}} \\
3 & (10y, 6m) & (0.907, -0.907) &  & 0.784 & 0.501 & -0.305 & -0.459 & 0.991
\\
6 & (10y, 3m) & (1.240, -1.240) &  & 0.870 & 0.582 & -0.266 & -0.461 & 1.000
\\
9 & (10y, 3m) & (1.606, -1.606) &  & 0.917 & 0.715 & -0.232 & -0.416 & 1.000
\\
12 & (20y, 3m) & (1.268, -1.268) &  & 0.903 & 0.802 & -0.253 & -0.368 & 0.965
\\
15 & (20y, 3m) & (1.023, -1.023) &  & 0.859 & 0.906 & -0.282 & -0.298 & 0.980
\\
18 & (20y, 6m) & (0.825, -0.825) &  & 0.809 & 0.966 & -0.309 & -0.280 & 0.992
\\
21 & (20y, 1y) & (0.750, -0.750) &  & 0.753 & 0.985 & -0.334 & -0.281 & 1.012
\\
24 & (20y, 1y) & (0.547, -0.547) &  & 0.677 & 0.988 & -0.355 & -0.294 & 1.029
\\
\multicolumn{9}{l}{\textit{Panel C. Generalized spread of the conventional pair}}
\\
3 & (10y, 3m) & (0.399, -0.724) &  & 0.860 & 0.571 & -0.280 & -0.562 & 0.894
\\
6 & (10y, 3m) & (0.839, -1.142) &  & 0.926 & 0.637 & -0.244 & -0.551 & 0.914
\\
9 & (10y, 3m) & (1.217, -1.469) &  & 0.946 & 0.747 & -0.220 & -0.480 & 0.939
\\
12 & (10y, 3m) & (1.105, -1.290) &  & 0.926 & 0.838 & -0.244 & -0.377 & 0.957
\\
15 & (10y, 3m) & (0.864, -0.995) &  & 0.863 & 0.924 & -0.283 & -0.309 & 0.970
\\
18 & (10y, 3m) & (0.628, -0.746) &  & 0.799 & 0.969 & -0.315 & -0.296 & 0.977
\\
21 & (10y, 3m) & (0.412, -0.512) &  & 0.714 & 0.975 & -0.345 & -0.308 & 0.985
\\
24 & (10y, 3m) & (0.243, -0.315) &  & 0.634 & 0.977 & -0.367 & -0.329 & 0.994
\\
\multicolumn{9}{l}{\textit{Panel D. Simple spread of the conventional pair}} \\
3 & (10y, 3m) & (0.811, -0.811) &  & 0.775 & 0.478 & -0.313 & -0.450 & 1.000
\\
6 & (10y, 3m) & (1.240, -1.240) &  & 0.870 & 0.582 & -0.266 & -0.461 & 1.000
\\
9 & (10y, 3m) & (1.606, -1.606) &  & 0.917 & 0.715 & -0.232 & -0.416 & 1.000
\\
12 & (10y, 3m) & (1.407, -1.407) &  & 0.905 & 0.823 & -0.251 & -0.333 & 1.000
\\
15 & (10y, 3m) & (1.077, -1.077) &  & 0.851 & 0.920 & -0.287 & -0.279 & 1.000
\\
18 & (10y, 3m) & (0.811, -0.811) &  & 0.781 & 0.971 & -0.319 & -0.272 & 1.000
\\
21 & (10y, 3m) & (0.559, -0.559) &  & 0.695 & 0.979 & -0.348 & -0.293 & 1.000
\\
24 & (10y, 3m) & (0.342, -0.342) &  & 0.617 & 0.978 & -0.369 & -0.322 & 1.000
\\ \hline
\end{tabular}
\end{center}
\end{table}

\begin{table}[p]
\caption{\textbf{Pair and coefficient selection from the training
period, 1961--2015} Panel A presents the results for the best pair and coefficients
selected by machine leaning for several forecasting horizons. Panel B
presents the recession prediction performance for the simple term spread of
the pair selected in Panel A. Panel C (D) presents the performance for the
generalized (simple) term spread of the conventional 10-year and three-month
pair. $\protect\lambda$ is the strength of the $L_1$ penalty, $\text{AUC}_\text{train}$ ($\text{AUC}_\text{test}$) is the area under the ROC in the training (test) period, log L (log PPL) is the average log likelihood (PPL) in the
training (test) period, and the EBF is the ratio of the
PPL of an alternative model to that of the benchmark model. The longest
maturity in the sample is 20 years.}
\label{tab:MainResult15}
\begin{center} \small
\begin{tabular}{ccccccccc}
\hline\hline
Horizon & Pair & $\boldsymbol{\beta}$ & $\lambda$ & $\text{AUC}_\text{train}$
& $\text{AUC}_\text{test}$ & log L & log PPL & EBF \\ \hline
\multicolumn{9}{l}{\textit{Panel A. Generalized spread of the ML pair}} \\
3 & (3y, 6m) & (0.398, -0.594) & 1.866 & 0.744 & 0.596 & -0.328 & -0.309 &
0.977 \\
6 & (10y, 3m) & (0.302, -0.494) & 4.595 & 0.799 & 0.832 & -0.313 & -0.282 &
0.967 \\
9 & (7y, 3m) & (0.923, -1.074) & 1.414 & 0.867 & 0.988 & -0.274 & -0.211 &
0.980 \\
12 & (7y, 3m) & (1.033, -1.147) & 1.231 & 0.883 & 1.000 & -0.273 & -0.199 &
0.987 \\
15 & (20y, 3m) & (0.609, -0.683) & 4.000 & 0.858 & 1.000 & -0.297 & -0.233 &
0.969 \\
18 & (20y, 3m) & (0.481, -0.556) & 4.925 & 0.832 & 0.996 & -0.315 & -0.253 &
0.984 \\
21 & (20y, 6m) & (0.509, -0.539) & 3.249 & 0.793 & 1.000 & -0.331 & -0.255 &
1.002 \\
24 & (20y, 1y) & (0.631, -0.611) & 1.625 & 0.747 & 0.996 & -0.344 & -0.249 &
1.023 \\
\multicolumn{9}{l}{\textit{Panel B. Simple spread of the ML pair}} \\
3 & (3y, 6m) & (1.102, -1.102) &  & 0.719 & 0.812 & -0.340 & -0.288 & 0.998
\\
6 & (10y, 3m) & (0.883, -0.883) &  & 0.791 & 0.902 & -0.314 & -0.248 & 1.000
\\
9 & (7y, 3m) & (1.337, -1.337) &  & 0.859 & 0.996 & -0.278 & -0.185 & 1.006
\\
12 & (7y, 3m) & (1.380, -1.380) &  & 0.876 & 1.000 & -0.274 & -0.177 & 1.009
\\
15 & (20y, 3m) & (0.993, -0.993) &  & 0.855 & 1.000 & -0.290 & -0.208 & 0.994
\\
18 & (20y, 3m) & (0.874, -0.874) &  & 0.828 & 0.988 & -0.306 & -0.232 & 1.004
\\
21 & (20y, 6m) & (0.725, -0.725) &  & 0.789 & 1.000 & -0.328 & -0.244 & 1.014
\\
24 & (20y, 1y) & (0.698, -0.698) &  & 0.748 & 0.996 & -0.344 & -0.244 & 1.028
\\
\multicolumn{9}{l}{\textit{Panel C. Generalized spread of the conventional pair}}
\\
3 & (10y, 3m) & (0.291, -0.481) &  & 0.735 & 0.660 & -0.334 & -0.302 & 0.984
\\
6 & (10y, 3m) & (0.632, -0.795) &  & 0.809 & 0.864 & -0.306 & -0.260 & 0.988
\\
9 & (10y, 3m) & (0.984, -1.108) &  & 0.858 & 0.988 & -0.276 & -0.206 & 0.985
\\
12 & (10y, 3m) & (1.075, -1.159) &  & 0.875 & 1.000 & -0.276 & -0.197 & 0.989
\\
15 & (10y, 3m) & (1.022, -1.071) &  & 0.859 & 1.000 & -0.290 & -0.208 & 0.994
\\
18 & (10y, 3m) & (0.859, -0.910) &  & 0.828 & 0.992 & -0.309 & -0.240 & 0.996
\\
21 & (10y, 3m) & (0.648, -0.696) &  & 0.772 & 1.000 & -0.336 & -0.259 & 0.998
\\
24 & (10y, 3m) & (0.484, -0.518) &  & 0.714 & 0.996 & -0.359 & -0.271 & 1.000
\\
\multicolumn{9}{l}{\textit{Panel D. Simple spread of the conventional pair}} \\
3 & (10y, 3m) & (0.576, -0.576) &  & 0.699 & 0.814 & -0.348 & -0.286 & 1.000
\\
6 & (10y, 3m) & (0.883, -0.883) &  & 0.791 & 0.902 & -0.314 & -0.248 & 1.000
\\
9 & (10y, 3m) & (1.193, -1.193) &  & 0.854 & 0.996 & -0.280 & -0.191 & 1.000
\\
12 & (10y, 3m) & (1.230, -1.230) &  & 0.872 & 1.000 & -0.277 & -0.186 & 1.000
\\
15 & (10y, 3m) & (1.120, -1.120) &  & 0.857 & 1.000 & -0.290 & -0.202 & 1.000
\\
18 & (10y, 3m) & (0.953, -0.953) &  & 0.821 & 0.992 & -0.310 & -0.237 & 1.000
\\
21 & (10y, 3m) & (0.733, -0.733) &  & 0.764 & 1.000 & -0.337 & -0.258 & 1.000
\\
24 & (10y, 3m) & (0.542, -0.542) &  & 0.710 & 0.984 & -0.359 & -0.271 & 1.000
\\ \hline
\end{tabular}
\end{center}
\end{table}

Note that if the training period extends to 2005 or 2015,
there are only one or two recession events in the out-of-sample data set. Thus, the
prediction that recessions will be absent would be correct almost all the
time. As a result, the AUC becomes close to one, particularly for long
forecasting horizons, and may not work effectively as a valid measure of
prediction performance.

\subsubsection{Imbalanced Classification}

The forecasting of recessions has a typical \textit{imbalanced} classification
problem in the sense that the period of a recession ($y_{t}=1$) forms only
a small fraction of the whole sample period. In such problems, the trained
models are heavily skewed to the majority class (i.e., non-recession), and
the prediction on the minority class (i.e., recession) is very poor. An ML practice that avoids this issue applies a weight
that is inversely proportional to the frequency of each class, which equalizes the importance
of the two classes. If the ratio of the recession in the
training period is $r$, the log likelihood is modified to
\begin{equation}
\log L(\beta_{0},\boldsymbol{\beta})=\sum_{t+k \in \mathcal{T}}w_{t+k}\left(
y_{t+k}\ln (\hat{y}_{t+k})+(1-y_{t+k})\ln (1-\hat{y}_{t+k})\right) ,
\label{eq:loglh_weight}
\end{equation}
where
\begin{equation*}
w_{t}=
\begin{cases}
\displaystyle\;\frac{1}{2r} & \text{{if }}y_{t}=1 \\
\displaystyle\frac{1}{2(1-r)} & \text{{if }}y_{t}=0
\end{cases}
.
\end{equation*}
This is equivalent to oversampling the recession observations $(1-r)/r$
times. Note that the original log likelihood, equation~\eqref{eq:loglh}, is
recovered when the recession and non-recession periods are equally balanced
as $r=1/2$.\footnote{
For the implementation, we use the ``balanced'' option for the \texttt{
class\_weight} parameter.} Here, the ratio $r$ is understood as a model
parameter inferred from the training period. As such, the same $r$ from the
training period should be used for the log likelihood over the test period
(i.e., log PPL).

\begin{table}[p]
\caption{\textbf{Pair and coefficient selection with recession oversampling} The results in this table are obtained
using the weighted regression in equation (\ref{eq:loglh_weight}), which has an
effect of oversampling one recession observation six ($=(1-r)/r$) times from
the recession ratio, $r=0.14$, in the training period, 1961--1995. Panel A presents the
results of the best pair and coefficients selected by machine leaning for
several forecasting horizons. Panel B presents the recession prediction
performance for the simple term spread of the pair selected in Panel A.
Panel C (D) presents the performance for the generalized (simple) term
spread of the conventional 10-year and three-month pair. $\protect\lambda$ is
the strength of the $L_1$ penalty, $\text{AUC}_\text{train}$ ($\text{AUC}_\text{test}$) is the area under the ROC
in the training (test) period, log L (log PPL) is the average log
 likelihood (PPL) in the training (test) period, and
the EBF is the ratio of
the PPL of an alternative model to that of the benchmark model.
The longest maturity in the sample is
20 years.}
\label{tab:MainResult95_weighted}
\begin{center} \small
\begin{tabular}{ccccccccc}
\hline\hline
Horizon & Pair & $\boldsymbol{\beta}$ & $\lambda$ & $\text{AUC}_\text{train}$
& $\text{AUC}_\text{test}$ & log L & log PPL & EBF \\ \hline
\multicolumn{9}{l}{\textit{Panel A. Generalized spread of the ML pair}} \\
3 & (20y, 3m) & (0.231, -0.691) & 7.464 & 0.889 & 0.516 & -0.448 & -1.259 &
0.615 \\
6 & (20y, 3m) & (0.576, -1.037) & 6.063 & 0.938 & 0.626 & -0.370 & -1.232 &
0.618 \\
9 & (20y, 3m) & (0.440, -0.800) & 10.556 & 0.932 & 0.745 & -0.402 & -0.822 &
0.809 \\
12 & (20y, 3m) & (0.361, -0.630) & 12.126 & 0.901 & 0.807 & -0.465 & -0.632
& 0.877 \\
15 & (20y, 3m) & (0.622, -0.799) & 5.278 & 0.864 & 0.879 & -0.488 & -0.519 &
0.927 \\
18 & (20y, 6m) & (0.720, -0.814) & 1.866 & 0.815 & 0.916 & -0.538 & -0.467 &
0.997 \\
21 & (20y, 1y) & (0.708, -0.750) & 1.414 & 0.753 & 0.915 & -0.591 & -0.473 &
1.052 \\
24 & (20y, 1y) & (0.413, -0.435) & 2.639 & 0.671 & 0.895 & -0.642 & -0.531 &
1.055 \\
\multicolumn{9}{l}{\textit{Panel B. Simple spread of the ML pair}} \\
3 & (20y, 3m) & (0.827, -0.827) &  & 0.794 & 0.503 & -0.550 & -0.859 & 0.917
\\
6 & (20y, 3m) & (1.208, -1.208) &  & 0.871 & 0.640 & -0.457 & -0.866 & 0.891
\\
9 & (20y, 3m) & (1.384, -1.384) &  & 0.896 & 0.771 & -0.417 & -0.703 & 0.911
\\
12 & (20y, 3m) & (1.372, -1.372) &  & 0.886 & 0.844 & -0.432 & -0.562 & 0.940
\\
15 & (20y, 3m) & (1.102, -1.102) &  & 0.850 & 0.898 & -0.487 & -0.454 & 0.990
\\
18 & (20y, 6m) & (0.922, -0.922) &  & 0.801 & 0.924 & -0.540 & -0.436 & 1.029
\\
21 & (20y, 1y) & (0.846, -0.846) &  & 0.746 & 0.920 & -0.590 & -0.456 & 1.071
\\
24 & (20y, 1y) & (0.568, -0.568) &  & 0.669 & 0.904 & -0.639 & -0.501 & 1.087
\\
\multicolumn{9}{l}{\textit{Panel C. Generalized spread of the conventional pair}}
\\
3 & (10y, 3m) & (0.642, -1.175) &  & 0.902 & 0.519 & -0.426 & -1.519 & 0.474
\\
6 & (10y, 3m) & (1.102, -1.655) &  & 0.940 & 0.631 & -0.346 & -1.443 & 0.501
\\
9 & (10y, 3m) & (1.354, -1.839) &  & 0.941 & 0.764 & -0.332 & -1.035 & 0.654
\\
12 & (10y, 3m) & (1.296, -1.599) &  & 0.916 & 0.832 & -0.387 & -0.655 & 0.857
\\
15 & (10y, 3m) & (0.979, -1.149) &  & 0.859 & 0.877 & -0.480 & -0.482 & 0.963
\\
18 & (10y, 3m) & (0.735, -0.845) &  & 0.796 & 0.893 & -0.554 & -0.467 & 0.997
\\
21 & (10y, 3m) & (0.478, -0.547) &  & 0.709 & 0.883 & -0.622 & -0.516 & 1.008
\\
24 & (10y, 3m) & (0.269, -0.310) &  & 0.616 & 0.880 & -0.668 & -0.576 & 1.009
\\
\multicolumn{9}{l}{\textit{Panel D. Simple spread of the conventional pair}} \\
3 & (10y, 3m) & (0.850, -0.850) &  & 0.794 & 0.522 & -0.555 & -0.772 & 1.000
\\
6 & (10y, 3m) & (1.270, -1.270) &  & 0.870 & 0.653 & -0.458 & -0.750 & 1.000
\\
9 & (10y, 3m) & (1.511, -1.511) &  & 0.901 & 0.779 & -0.409 & -0.610 & 1.000
\\
12 & (10y, 3m) & (1.513, -1.513) &  & 0.894 & 0.846 & -0.421 & -0.500 & 1.000
\\
15 & (10y, 3m) & (1.141, -1.141) &  & 0.850 & 0.889 & -0.492 & -0.444 & 1.000
\\
18 & (10y, 3m) & (0.851, -0.851) &  & 0.783 & 0.901 & -0.560 & -0.464 & 1.000
\\
21 & (10y, 3m) & (0.559, -0.559) &  & 0.697 & 0.890 & -0.625 & -0.524 & 1.000
\\
24 & (10y, 3m) & (0.320, -0.320) &  & 0.610 & 0.892 & -0.670 & -0.585 & 1.000
\\ \hline
\end{tabular}
\end{center}
\end{table}

\begin{table}[p]
\caption{\textbf{Pair and coefficient selection with leading indicator included} The results in this table are
obtained by including the US leading indicator as a default
variable, in addition to the ML or conventional pair. Panel A presents the
results of the best pair and coefficients selected by machine leaning for
several forecasting horizons. Panel B presents the recession prediction
performance for the simple term spread of the pair selected in Panel A.
Panel C (D) presents the performance for the generalized (simple) term
spread of the conventional 10-year and three-month pair. $\protect\beta_\text{LI}
$ is the coefficient of the leading indicator, $\protect\lambda$ is the
strength of the $L_1$ penalty, $\text{AUC}_\text{train}$ ($\text{AUC}_\text{
test}$) is the area under the ROC in the
training (test) period, log L (log PPL) is the average log likelihood (PPL) in the training (test) period, and the EBF is the ratio of the
PPL of an alternative model to that of the benchmark model.
The training period is from June 1982 to December
1995. The longest maturity in the sample is 30 years.}
\label{tab:MainResult95_2_USSLIND}
\begin{center} \footnotesize
\begin{tabular}{cccccccccc}
\hline\hline
Horizon & Pair & $\boldsymbol{\beta}$ & $\beta_\text{LI}$ & $\lambda $ & $
\text{AUC}_\text{train}$ & $\text{AUC}_\text{test}$ & log L & log PPL
& EBF \\ \hline
\multicolumn{10}{l}{\textit{Panel A. Generalized spread of the ML pair}} \\
3 & (30y, 3m) & (0.632, -1.426) & 2.923 & 0.354 & 0.983 & 0.878 & -0.089 &
-0.417 & 0.797 \\
6 & (30y, 3m) & (0.364, -0.627) & 1.020 & 1.414 & 0.931 & 0.859 & -0.165 &
-0.311 & 0.934 \\
9 & (30y, 3m) & (1.005, -1.038) & 0.350 & 0.812 & 0.926 & 0.823 & -0.155 &
-0.274 & 0.988 \\
12 & (30y, 3m) & (1.091, -0.831) & 0.013 & 1.516 & 0.977 & 0.845 & -0.126 &
-0.280 & 0.996 \\
15 & (30y, 3m) & (1.752, -1.275) & 0.481 & 1.000 & 1.000 & 0.912 & -0.090 &
-0.308 & 0.927 \\
18 & (30y, 3m) & (1.375, -1.216) & -0.150 & 1.072 & 0.986 & 0.915 & -0.111 &
-0.219 & 1.013 \\
21 & (30y, 3m) & (1.130, -1.109) & -0.854 & 1.231 & 0.980 & 0.881 & -0.126 &
-0.248 & 1.005 \\
24 & (30y, 6m) & (0.688, -0.606) & -0.862 & 1.414 & 0.924 & 0.888 & -0.167 &
-0.265 & 0.983 \\
\multicolumn{10}{l}{\textit{Panel B. Simple spread of the ML pair}} \\
3 & (30y, 3m) & (0.886, -0.886) & 2.427 &  & 0.979 & 0.927 & -0.115 & -0.194
& 0.997 \\
6 & (30y, 3m) & (1.326, -1.326) & 1.262 &  & 0.939 & 0.848 & -0.150 & -0.268
& 0.975 \\
9 & (30y, 3m) & (1.510, -1.510) & 0.290 &  & 0.929 & 0.809 & -0.146 & -0.290
& 0.972 \\
12 & (30y, 3m) & (2.008, -2.008) & -0.499 &  & 0.964 & 0.821 & -0.101 &
-0.300 & 0.977 \\
15 & (30y, 3m) & (2.307, -2.307) & -0.086 &  & 0.988 & 0.894 & -0.084 &
-0.246 & 0.986 \\
18 & (30y, 3m) & (2.335, -2.335) & -0.505 &  & 0.990 & 0.903 & -0.084 &
-0.235 & 0.998 \\
21 & (30y, 3m) & (2.090, -2.090) & -1.282 &  & 0.981 & 0.887 & -0.096 &
-0.250 & 1.003 \\
24 & (30y, 6m) & (1.558, -1.558) & -1.291 &  & 0.916 & 0.897 & -0.139 &
-0.244 & 1.003 \\
\multicolumn{10}{l}{\textit{Panel C. Generalized spread of the conventional pair}
} \\
3 & (10y, 3m) & (0.438, -1.134) & 2.172 &  & 0.984 & 0.877 & -0.097 & -0.381
& 0.826 \\
6 & (10y, 3m) & (0.926, -1.278) & 0.968 &  & 0.938 & 0.792 & -0.148 & -0.350
& 0.898 \\
9 & (10y, 3m) & (1.300, -1.434) & 0.267 &  & 0.928 & 0.813 & -0.148 & -0.283
& 0.979 \\
12 & (10y, 3m) & (2.013, -1.708) & -0.132 &  & 0.981 & 0.832 & -0.096 &
-0.323 & 0.954 \\
15 & (10y, 3m) & (2.273, -1.887) & 0.418 &  & 1.000 & 0.894 & -0.073 & -0.314
& 0.921 \\
18 & (10y, 3m) & (1.877, -1.838) & -0.224 &  & 0.984 & 0.894 & -0.097 &
-0.228 & 1.005 \\
21 & (10y, 3m) & (1.630, -1.773) & -0.997 &  & 0.976 & 0.861 & -0.112 &
-0.271 & 0.983 \\
24 & (10y, 3m) & (1.148, -1.188) & -1.057 &  & 0.895 & 0.878 & -0.157 &
-0.259 & 0.989 \\
\multicolumn{10}{l}{\textit{Panel D. Simple spread of the conventional pair}} \\
3 & (10y, 3m) & (0.779, -0.779) & 2.453 &  & 0.974 & 0.939 & -0.123 & -0.191
& 1.000 \\
6 & (10y, 3m) & (1.310, -1.310) & 1.345 &  & 0.934 & 0.893 & -0.157 & -0.242
& 1.000 \\
9 & (10y, 3m) & (1.615, -1.615) & 0.394 &  & 0.930 & 0.821 & -0.146 & -0.261
& 1.000 \\
12 & (10y, 3m) & (2.222, -2.222) & -0.426 &  & 0.973 & 0.825 & -0.093 &
-0.276 & 1.000 \\
15 & (10y, 3m) & (2.589, -2.589) & 0.057 &  & 0.998 & 0.892 & -0.071 & -0.232
& 1.000 \\
18 & (10y, 3m) & (2.371, -2.371) & -0.370 &  & 0.984 & 0.893 & -0.087 &
-0.233 & 1.000 \\
21 & (10y, 3m) & (2.061, -2.061) & -1.179 &  & 0.969 & 0.869 & -0.105 &
-0.253 & 1.000 \\
24 & (10y, 3m) & (1.366, -1.366) & -1.202 &  & 0.895 & 0.884 & -0.153 &
-0.247 & 1.000 \\ \hline
\end{tabular}
\end{center}
\end{table}

Table~\ref{tab:MainResult95_weighted} reports the result for the case when
the training period runs until 1995. Because the ratio of the recession during the
training period is approximately $r=0.14$, the weighted regression is
equivalent to oversampling the recessions six times. Under this test, our
main findings remain unchanged. In particular, the superior performance of the
simple term spread (Panels B and D) over the generalized term spread (Panels
A and C) is more pronounced than in Table~\ref{tab:MainResult95}.

\subsubsection{Control Variables}

Finally, we repeat the analyses with additional recession predictors in
order to account for the missing variable problem. Specifically, we include
the leading business cycle indicator and ensure that the variable is always
selected in the ML approach by excluding its coefficient in
the $L_{1}$ penalty term.\footnote{
The US leading business cycle indicator is available from the Saint Louis
Fed website.} We also consider the 30-year Treasury yield, which makes our
training period start from 1982. This examination shows whether
the yield pair from ML has any additional predictive
ability beyond that which the leading indicator already explains. It is also worth
checking whether a pair with one short-term and one long-term yield is
still chosen by the ML method, even when the leading indicator is
already controlled.\footnote{
Because the 10-year--three-month spread is one component in the US
leading indicator, we are ex ante agnostic about whether the pairs from
ML are again composed of one short- and one long-term yield.}

Table~\ref{tab:MainResult95_2_USSLIND} shows that even when the leading
indicator is included as a default variable, the pair choice from machine
learning is rarely affected: for most forecasting horizons, one long-term
and one short-term yield are still chosen, and the out-of-sample prediction
performance is not improved significantly over that of the conventional
10-year--three-month pair. The weight for model averaging is again near 0.5, indicating
that the performance of the pair from ML is almost equal to
that of the conventional pair.\footnote{In an unreported analysis, we also examine alternative control variables such as the CBOE Volatility Index and the Pastor-Stambaugh liquidity factor, and confirm the robustness of the findings. With the alternative control variable included, the largest EBF is only 1.069, meaning that the gain from the generalized spread of the ML pair is marginal. We thank the anonymous referee for a helpful suggestion.}

\section{Conclusion}

\label{sec:conc} \noindent The ten-year and three-month term spread is
widely accepted in estimating predictive recession probabilities. The
contribution of our study is to provide a justification for using the
conventional term spread to predict such probabilities. To this end,
we formally and comprehensively test whether this prediction ability can be
improved. Using the ML approach, we identify the optimal maturity pair, allowing for separate
regression effects of short- and long-term interest rates. According to
our empirical exercise, relaxing the restrictions on the maturity pair and
coefficients does not improve the out-of-sample prediction ability of the yield spread. This is due to the dominant estimation risk, which is in line with the result of \citet{puglia2020ml}.
Our finding is not sensitive to the
forecasting horizon, sample period, oversampling problem, or control
variable.

\bigskip
\bigskip
\singlespacing
\bibliography{YieldCurve}

\begin{thebibliography}{29}
\expandafter\ifx\csname natexlab\endcsname\relax\def\natexlab#1{#1}\fi
\providecommand{\url}[1]{\texttt{#1}}
\providecommand{\href}[2]{#2}
\providecommand{\path}[1]{#1}
\providecommand{\DOIprefix}{doi:}
\providecommand{\ArXivprefix}{arXiv:}
\providecommand{\URLprefix}{URL: }
\providecommand{\Pubmedprefix}{pmid:}
\providecommand{\doi}[1]{\href{http://dx.doi.org/#1}{\path{#1}}}
\providecommand{\Pubmed}[1]{\href{pmid:#1}{\path{#1}}}
\providecommand{\bibinfo}[2]{#2}
\ifx\xfnm\relax \def\xfnm[#1]{\unskip,\space#1}\fi
\bibitem[{Bauer and Mertens(2018a)}]{bauer2018economic}
\bibinfo{author}{Bauer, M.D.}, \bibinfo{author}{Mertens, T.M.},
  \bibinfo{year}{2018}a.
\newblock \bibinfo{title}{Economic {{Forecasts}} with the {{Yield Curve}}}.
\newblock \bibinfo{type}{{{FRBSF Economic Letter}}} \bibinfo{number}{2018-07}.
  {Federal Reserve Bank of San Francisco}.
\newblock \URLprefix
  \url{https://www.frbsf.org/economic-research/publications/economic-letter/2018/march/economic-forecasts-with-yield-curve/}.
\bibitem[{Bauer and Mertens(2018b)}]{bauer2018information}
\bibinfo{author}{Bauer, M.D.}, \bibinfo{author}{Mertens, T.M.},
  \bibinfo{year}{2018}b.
\newblock \bibinfo{title}{Information in the {{Yield Curve}} about {{Future
  Recessions}}}.
\newblock \bibinfo{type}{{{FRBSF Economic Letter}}} \bibinfo{number}{2018-20}.
  {Federal Reserve Bank of San Francisco}.
\newblock \URLprefix
  \url{https://www.frbsf.org/economic-research/publications/economic-letter/2018/august/information-in-yield-curve-about-future-recessions/}.
\bibitem[{Berge and Jord{\`a}(2011)}]{Berge2011}
\bibinfo{author}{Berge, T.J.}, \bibinfo{author}{Jord{\`a}, {\`O}.},
  \bibinfo{year}{2011}.
\newblock \bibinfo{title}{Evaluating the {{Classification}} of {{Economic
  Activity}} into {{Recessions}} and {{Expansions}}}.
\newblock \bibinfo{journal}{American Economic Journal: Macroeconomics}
  \bibinfo{volume}{3}, \bibinfo{pages}{246--277}.
\newblock \DOIprefix\doi{10.1257/mac.3.2.246}.
\bibitem[{Bradley(1997)}]{bradley1997use}
\bibinfo{author}{Bradley, A.P.}, \bibinfo{year}{1997}.
\newblock \bibinfo{title}{The use of the area under the {{ROC}} curve in the
  evaluation of machine learning algorithms}.
\newblock \bibinfo{journal}{Pattern Recognition} \bibinfo{volume}{30},
  \bibinfo{pages}{1145--1159}.
\newblock \DOIprefix\doi{10.1016/S0031-3203(96)00142-2}.
\bibitem[{DeMiguel et~al.(2009)DeMiguel, Garlappi and
  Uppal}]{demiguel2009naive}
\bibinfo{author}{DeMiguel, V.}, \bibinfo{author}{Garlappi, L.},
  \bibinfo{author}{Uppal, R.}, \bibinfo{year}{2009}.
\newblock \bibinfo{title}{Optimal {{Versus Naive Diversification}}: {{How
  Inefficient}} is the 1/{{N Portfolio Strategy}}?}
\newblock \bibinfo{journal}{The Review of Financial Studies}
  \bibinfo{volume}{22}, \bibinfo{pages}{1915--1953}.
\newblock \DOIprefix\doi{10.1093/rfs/hhm075}.
\bibitem[{D{\"o}pke et~al.(2017)D{\"o}pke, Fritsche and
  Pierdzioch}]{dopke2017predicting}
\bibinfo{author}{D{\"o}pke, J.}, \bibinfo{author}{Fritsche, U.},
  \bibinfo{author}{Pierdzioch, C.}, \bibinfo{year}{2017}.
\newblock \bibinfo{title}{Predicting recessions with boosted regression trees}.
\newblock \bibinfo{journal}{International Journal of Forecasting}
  \bibinfo{volume}{33}, \bibinfo{pages}{745--759}.
\newblock \DOIprefix\doi{10.1016/j.ijforecast.2017.02.003}.
\bibitem[{Engstrom and Sharpe(2018)}]{engstrom2018near}
\bibinfo{author}{Engstrom, E.C.}, \bibinfo{author}{Sharpe, S.A.},
  \bibinfo{year}{2018}.
\newblock \bibinfo{title}{The {{Near-Term Forward Yield Spread}} as a {{Leading
  Indicator}}: {{A Less Distorted Mirror}}}.
\newblock \bibinfo{type}{Finance and {{Economics Discussion Series}}}
  \bibinfo{number}{2018-055}. {Washington: Board of Governors of the Federal
  Reserve System}.
\newblock \DOIprefix\doi{10.17016/FEDS.2018.055r1}.
\bibitem[{Ergungor(2016)}]{ergungor2016recession}
\bibinfo{author}{Ergungor, O.E.}, \bibinfo{year}{2016}.
\newblock \bibinfo{title}{Recession Probabilities}.
\newblock \bibinfo{type}{Economic {{Commentary}}} \bibinfo{number}{2016-09}.
  {Federal Reserve Bank of Cleveland}.
\newblock \URLprefix
  \url{https://www.clevelandfed.org/en/newsroom-and-events/publications/economic-commentary/2016-economic-commentaries/ec-201609-recession-probabilities.aspx}.
\bibitem[{Estrella(2005)}]{estrella2005yield}
\bibinfo{author}{Estrella, A.}, \bibinfo{year}{2005}.
\newblock \bibinfo{title}{The {{Yield Curve}} as a {{Leading Indicator}}}.
\newblock \bibinfo{type}{Technical Report}. {Federal Reserve Bank of New York}.
\newblock \URLprefix
  \url{https://www.newyorkfed.org/research/capital_markets/ycfaq.html}.
\bibitem[{Estrella and Hardouvelis(1991)}]{estrella1991term}
\bibinfo{author}{Estrella, A.}, \bibinfo{author}{Hardouvelis, G.A.},
  \bibinfo{year}{1991}.
\newblock \bibinfo{title}{The {{Term Structure}} as a {{Predictor}} of {{Real
  Economic Activity}}}.
\newblock \bibinfo{journal}{The Journal of Finance} \bibinfo{volume}{46},
  \bibinfo{pages}{555--576}.
\newblock \DOIprefix\doi{10.1111/j.1540-6261.1991.tb02674.x}.
\bibitem[{Estrella and Trubin(2006)}]{estrella2006yield}
\bibinfo{author}{Estrella, A.}, \bibinfo{author}{Trubin, M.},
  \bibinfo{year}{2006}.
\newblock \bibinfo{title}{The {{Yield Curve}} as a {{Leading Indicator}}:
  {{Some Practical Issues}}}.
\newblock \bibinfo{type}{Current {{Issues}} in {{Economics}} and {{Finance}}}
  \bibinfo{number}{July/August 2006 Volume 12, Number 5}. {Federal Reserve Bank
  of New York}.
\newblock \URLprefix
  \url{https://www.newyorkfed.org/research/current_issues/ci12-5.html}.
\bibitem[{Gogas et~al.(2015)Gogas, Papadimitriou, Matthaiou and
  Chrysanthidou}]{gogas2015yield}
\bibinfo{author}{Gogas, P.}, \bibinfo{author}{Papadimitriou, T.},
  \bibinfo{author}{Matthaiou, M.}, \bibinfo{author}{Chrysanthidou, E.},
  \bibinfo{year}{2015}.
\newblock \bibinfo{title}{Yield {{Curve}} and {{Recession Forecasting}} in a
  {{Machine Learning Framework}}}.
\newblock \bibinfo{journal}{Computational Economics} \bibinfo{volume}{45},
  \bibinfo{pages}{635--645}.
\newblock \DOIprefix\doi{10.1007/s10614-014-9432-0}.
\bibitem[{Goyenko et~al.(2011)Goyenko, Subrahmanyam and Ukhov}]{Goyenko2011}
\bibinfo{author}{Goyenko, R.}, \bibinfo{author}{Subrahmanyam, A.},
  \bibinfo{author}{Ukhov, A.}, \bibinfo{year}{2011}.
\newblock \bibinfo{title}{The {{Term Structure}} of {{Bond Market Liquidity}}
  and {{Its Implications}} for {{Expected Bond Returns}}}.
\newblock \bibinfo{journal}{The Journal of Financial and Quantitative Analysis}
  \bibinfo{volume}{46}, \bibinfo{pages}{111--139}.
\newblock \URLprefix \url{https://www.jstor.org/stable/23018519}.
\bibitem[{G{\"u}rkaynak et~al.(2006)G{\"u}rkaynak, Sack and
  Wright}]{gurkaynak2006treasury}
\bibinfo{author}{G{\"u}rkaynak, R.S.}, \bibinfo{author}{Sack, B.},
  \bibinfo{author}{Wright, J.H.}, \bibinfo{year}{2006}.
\newblock \bibinfo{title}{The {{U}}.{{S}}. {{Treasury}} Yield Curve: 1961 to
  the Present}.
\newblock \bibinfo{type}{Finance and {{Economics Discussion Series}}}
  \bibinfo{number}{2006-28}. {Washington: Board of Governors of the Federal
  Reserve System}.
\newblock \URLprefix
  \url{https://www.federalreserve.gov/pubs/feds/2006/200628/200628abs.html}.
\bibitem[{G{\"u}rkaynak et~al.(2007)G{\"u}rkaynak, Sack and
  Wright}]{gurkaynak2007treasury}
\bibinfo{author}{G{\"u}rkaynak, R.S.}, \bibinfo{author}{Sack, B.},
  \bibinfo{author}{Wright, J.H.}, \bibinfo{year}{2007}.
\newblock \bibinfo{title}{The {{U}}.{{S}}. {{Treasury}} yield curve: 1961 to
  the present}.
\newblock \bibinfo{journal}{Journal of Monetary Economics}
  \bibinfo{volume}{54}, \bibinfo{pages}{2291--2304}.
\newblock \DOIprefix\doi{10.1016/j.jmoneco.2007.06.029}.
\bibitem[{Hall(2018)}]{hall2018machine}
\bibinfo{author}{Hall, A.S.}, \bibinfo{year}{2018}.
\newblock \bibinfo{title}{Machine {{Learning Approaches}} to {{Macroeconomic
  Forecasting}}}.
\newblock \bibinfo{type}{Economic {{Review}}} \bibinfo{number}{4th Quarter
  2018}. {Fedral Reserve Bank of Kansas City}.
\newblock \DOIprefix\doi{10.18651/ER/4q18SmalterHall}.
\bibitem[{Hastie et~al.(2009)Hastie, Tibshirani and Friedman}]{hastie2009ESL}
\bibinfo{author}{Hastie, T.}, \bibinfo{author}{Tibshirani, R.},
  \bibinfo{author}{Friedman, J.}, \bibinfo{year}{2009}.
\newblock \bibinfo{title}{The {{Elements}} of {{Statistical Learning}}: {{Data
  Mining}}, {{Inference}}, and {{Prediction}}, {{Second Edition}}}.
\newblock \bibinfo{edition}{2nd edition} ed., \bibinfo{address}{{New York,
  NY}}.
\newblock \URLprefix \url{https://web.stanford.edu/~hastie/ElemStatLearn/}.
\bibitem[{Johansson and Meldrum(2018)}]{Johansson2018}
\bibinfo{author}{Johansson, P.}, \bibinfo{author}{Meldrum, A.},
  \bibinfo{year}{2018}.
\newblock \bibinfo{title}{Predicting {{Recession Probabilities Using}} the
  {{Slope}} of the {{Yield Curve}}}.
\newblock \bibinfo{type}{{{FEDS Notes}}}. {Washington: Board of Governors of
  the Federal Reserve System}.
\newblock \DOIprefix\doi{10.17016/2380-7172.2146}.
\bibitem[{Kim and Swanson(2018)}]{kim2018mining}
\bibinfo{author}{Kim, H.H.}, \bibinfo{author}{Swanson, N.R.},
  \bibinfo{year}{2018}.
\newblock \bibinfo{title}{Mining big data using parsimonious factor, machine
  learning, variable selection and shrinkage methods}.
\newblock \bibinfo{journal}{International Journal of Forecasting}
  \bibinfo{volume}{34}, \bibinfo{pages}{339--354}.
\newblock \DOIprefix\doi{10.1016/j.ijforecast.2016.02.012}.
\bibitem[{Liu and Moench(2016)}]{liu2016what}
\bibinfo{author}{Liu, W.}, \bibinfo{author}{Moench, E.}, \bibinfo{year}{2016}.
\newblock \bibinfo{title}{What predicts {{US}} recessions?}
\newblock \bibinfo{journal}{International Journal of Forecasting}
  \bibinfo{volume}{32}, \bibinfo{pages}{1138--1150}.
\newblock \DOIprefix\doi{10.1016/j.ijforecast.2016.02.007}.
\bibitem[{Pedregosa et~al.(2011)Pedregosa, Varoquaux, Gramfort, Michel,
  Thirion, Grisel, Blondel, Prettenhofer, Weiss, Dubourg, Vanderplas, Passos,
  Cournapeau, Brucher, Perrot and Duchesnay}]{pedregosa2011sklearn}
\bibinfo{author}{Pedregosa, F.}, \bibinfo{author}{Varoquaux, G.},
  \bibinfo{author}{Gramfort, A.}, \bibinfo{author}{Michel, V.},
  \bibinfo{author}{Thirion, B.}, \bibinfo{author}{Grisel, O.},
  \bibinfo{author}{Blondel, M.}, \bibinfo{author}{Prettenhofer, P.},
  \bibinfo{author}{Weiss, R.}, \bibinfo{author}{Dubourg, V.},
  \bibinfo{author}{Vanderplas, J.}, \bibinfo{author}{Passos, A.},
  \bibinfo{author}{Cournapeau, D.}, \bibinfo{author}{Brucher, M.},
  \bibinfo{author}{Perrot, M.}, \bibinfo{author}{Duchesnay, {\'E}.},
  \bibinfo{year}{2011}.
\newblock \bibinfo{title}{Scikit-learn: {{Machine Learning}} in {{Python}}}.
\newblock \bibinfo{journal}{Journal of Machine Learning Research}
  \bibinfo{volume}{12}, \bibinfo{pages}{2825--2830}.
\newblock \URLprefix \url{http://jmlr.org/papers/v12/pedregosa11a.html}.
\bibitem[{Puglia and Tucker(2020)}]{puglia2020ml}
\bibinfo{author}{Puglia, M.}, \bibinfo{author}{Tucker, A.},
  \bibinfo{year}{2020}.
\newblock \bibinfo{title}{Machine {{Learning}}, the {{Treasury Yield Curve}}
  and {{Recession Forecasting}}}.
\newblock \bibinfo{type}{Finance and {{Economics Discussion Series}}}
  \bibinfo{number}{2020-038}. {Washington: Board of Governors of the Federal
  Reserve System}.
\newblock \DOIprefix\doi{10.17016/FEDS.2020.038}.
\bibitem[{Racicot et~al.(2007)Racicot, Th{\'e}oret and
  Co{\"e}n}]{racicot2007forecasting}
\bibinfo{author}{Racicot, F.{\'E}.}, \bibinfo{author}{Th{\'e}oret, R.},
  \bibinfo{author}{Co{\"e}n, A.}, \bibinfo{year}{2007}.
\newblock \bibinfo{title}{Forecasting {{UHF Financial Data}}: {{Realized
  Volatility}} versus {{UHF-GARCH Models}}}.
\newblock \bibinfo{journal}{International Advances in Economic Research}
  \bibinfo{volume}{13}, \bibinfo{pages}{243--244}.
\newblock \DOIprefix\doi{10.1007/s11294-007-9079-x}.
\bibitem[{Rudebusch and Williams(2009)}]{rudebusch2009forecasting}
\bibinfo{author}{Rudebusch, G.D.}, \bibinfo{author}{Williams, J.C.},
  \bibinfo{year}{2009}.
\newblock \bibinfo{title}{Forecasting {{Recessions}}: {{The Puzzle}} of the
  {{Enduring Power}} of the {{Yield Curve}}}.
\newblock \bibinfo{journal}{Journal of Business \& Economic Statistics}
  \bibinfo{volume}{27}, \bibinfo{pages}{492--503}.
\newblock \DOIprefix\doi{10.1198/jbes.2009.07213}.
\bibitem[{Stekler and Ye(2017)}]{stekler2017evaluating}
\bibinfo{author}{Stekler, H.O.}, \bibinfo{author}{Ye, T.},
  \bibinfo{year}{2017}.
\newblock \bibinfo{title}{Evaluating a leading indicator: An
  application\textemdash the term spread}.
\newblock \bibinfo{journal}{Empirical Economics} \bibinfo{volume}{53},
  \bibinfo{pages}{183--194}.
\newblock \DOIprefix\doi{10.1007/s00181-016-1200-7}.
\bibitem[{Stock and Watson(1989)}]{stock1989new}
\bibinfo{author}{Stock, J.H.}, \bibinfo{author}{Watson, M.W.},
  \bibinfo{year}{1989}.
\newblock \bibinfo{title}{New {{Indexes}} of {{Coincident}} and {{Leading
  Economic Indicators}}}.
\newblock \bibinfo{journal}{NBER Macroeconomics Annual} \bibinfo{volume}{4},
  \bibinfo{pages}{351--394}.
\newblock \DOIprefix\doi{10.1086/654119}.
\bibitem[{Tsang and Wu(2019)}]{tsang2019revisiting}
\bibinfo{author}{Tsang, E.}, \bibinfo{author}{Wu, M.}, \bibinfo{year}{2019}.
\newblock \bibinfo{title}{Revisiting {{US Recession Probability Models}}}.
\newblock \bibinfo{type}{Research {{Memorandum}}} \bibinfo{number}{03/2019}.
  {Hong Kong Monetary Authority}.
\newblock \URLprefix
  \url{https://www.hkma.gov.hk/media/eng/publication-and-research/research/research-memorandums/2019/RM03-2019.pdf}.
\bibitem[{Welch and Goyal(2008)}]{welch2008equitypred}
\bibinfo{author}{Welch, I.}, \bibinfo{author}{Goyal, A.}, \bibinfo{year}{2008}.
\newblock \bibinfo{title}{A {{Comprehensive Look}} at {{The Empirical
  Performance}} of {{Equity Premium Prediction}}}.
\newblock \bibinfo{journal}{The Review of Financial Studies}
  \bibinfo{volume}{21}, \bibinfo{pages}{1455--1508}.
\newblock \DOIprefix\doi{10.1093/rfs/hhm014}.
\bibitem[{Zou and Hastie(2005)}]{zou2005elastic}
\bibinfo{author}{Zou, H.}, \bibinfo{author}{Hastie, T.}, \bibinfo{year}{2005}.
\newblock \bibinfo{title}{Regularization and variable selection via the elastic
  net}.
\newblock \bibinfo{journal}{Journal of the Royal Statistical Society: Series B
  (Statistical Methodology)} \bibinfo{volume}{67}, \bibinfo{pages}{301--320}.
\newblock \DOIprefix\doi{10.1111/j.1467-9868.2005.00503.x}.

\end{thebibliography}

\end{document}